# *RSstitcher*: Merging 2D diffraction frames for Wide Range Reciprocal Space Maps with absorption correction and integration functions

Authors

**Xiaodong Wang[a]*, Michael W. M. Jones[abcd] and Adam Smith[e]**

[a]Central Analytical Research Facility, Queensland University of Technology (QUT), Level 6, P Block, Gardens Point Campus, Brisbane, Queensland, 4001, Australia

[b]Centre for Materials Science, Queensland University of Technology, Brisbane, Queensland, 4001, Australia

[c]Planetary Surface Exploration Group, Queensland University of Technology, Brisbane, Queensland, 4001, Australia

[d]School of Chemistry & Physics, Queensland University of Technology, Brisbane, Queensland, 4001, Australia

[e]eResearch, Queensland University of Technology, Brisbane, Queensland, 4001, Australia

Correspondence email: tony.wang@qut.edu.au

**Synopsis** *RSstitcher* (Reciprocal Space Stitcher) is an open-source python program converting 2D diffraction frames measured at different cradle tilt angles into Wide Range Reciprocal Space Maps (WRRSMs). It is so far the only WRRSMs generation tool that applied diffraction intensity correction due to sample self-absorption and can merge 2D frames measured in either symmetric scan mode or in ω-φ compensated Side Inclination Grazing Incident Diffraction mode.

**Abstract** Wide Range Reciprocal Space Mapping (WRRSM) is a technique that allows visualisation of the geometric relationships among multiple *hkl* spots in a whole reciprocal space map. However, commercial softwares for WRRSMs generation are associated with several issues or limitations, which are overcome by the currently reporting open-source python program *RSstitcher* (Reciprocal Space Stitcher). *RSstitcher* merges 2D scan frames formats supported by *FabIO* and enables WRRSM function on most laboratory X-ray diffractometers equipped with a goniometer cradle and a 2D detector of any sensor size. It is so far the only WRRSMs generation tool that applies diffraction intensity correction due to sample self-absorption, which enables quantitative analyses for WRRSMs including texture measurement and 1D data integration. The conversion equations used in the python program is explained geometrically, including novel WRRSM measurements in ω-φ compensated Side Inclination Grazing Incident Diffraction mode for thin film samples. The applications of *RSstitcher* for bulk and thin film samples are demonstrated using two common 2D X-ray diffraction systems.

## 1. Introduction

With the increasing demand for high-performance materials in semiconductor, optoelectronic, and energy applications, the need for advanced crystal structure characterization tools such as Wide Range Reciprocal Space Mapping (WRRSM) for bulks, thin films and devices has become more pronounced. The technique of WRRSM not only provides a rich dataset for direct visualisation of the sample's reciprocal space lattices and configurations but also enhances the interpretability of complex crystal texture that are often missed in conventional X-ray diffraction (XRD) measurements. In contrast to single crystal XRD which operates in transmission geometry for micron-sized single crystals, WRRSM mostly operates in reflection geometry for most bulk, thin film, and device samples on powder diffractometers requiring minimal sample preparation.

In X-ray crystallography, a reciprocal space spot corresponds to a family of equal-spacing parallel crystal planes in real space. A crystal lattice in real space is therefore able to be fully described by its reciprocal lattice. Reciprocal Space Mapping (RSM) is an XRD technique that requires highly parallel and monochromatic incident and diffracted X-ray beams to visualise the shape and vicinity of a particular reciprocal space *hkl* spot of epitaxial thin films and substrates in the goniometer's equatorial plane (Figure 1a). Conventionally, RSM enables the characterization of lattice parameters, strain-relaxation states, doping levels, and mosaicity of epitaxial layers by focusing on the analysis of relative locations of a specific reciprocal space spot from epitaxial layers and the substrate (Fewster, 1996; Ulyanenkova *et al.*, 2013; Poulsen *et al.*, 2018).

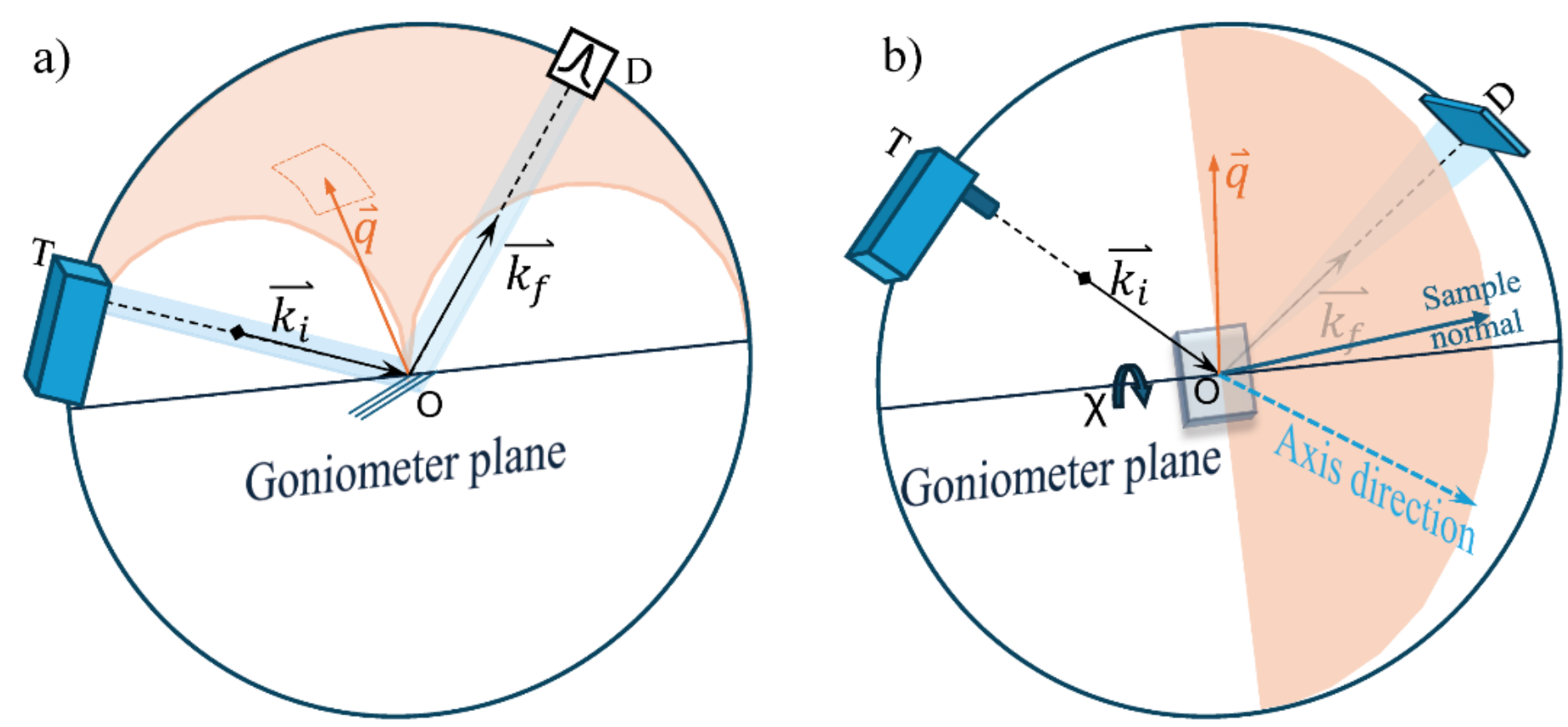

**Figure 1** **a)** Conventional RSM can theoretically access the pink region however practically only measure the vicinity (dotted box) of a particular reciprocal spot ($\vec{q} = \overrightarrow{k_f} - \overrightarrow{k_i}$) in the goniometer plane (TOD) using parallel X-ray beam and 0D detector; **b)** By tilting the sample surface along the cradle χ axis, WRRSM is able to measure the whole upper semicircle plane perpendicular to the goniometer plane (TOD) using point focus X-ray beam, a cradle and a 2D detector. T: X-ray Tube; D: Detector; O: goniometer centre; $\overrightarrow{k_i}$: incident X-ray wavevector; $\overrightarrow{k_f}$: diffract X-ray wavevector; $\vec{q}$: scattering vector. Orange area: the mappable reciprocal space.

In contrast, WRRSM significantly expands the coverage of measurable reciprocal space by tilting sample around the χ axis on a Eulerian cradle (Figure 1b). The mapped reciprocal space is perpendicular to the goniometer plane. When the cradle χ tilt range is limited to positive 0° to 90°, 2D scans in opposite azimuthal directions should be combined to cover the whole upper semicircle of the sample reciprocal space. Compared to RSM, WRRSM facilitates an overview and a global understanding of complex crystal orientations and the organisation of heterostructures. Since a wide reciprocal plane is measured, besides epitaxial thin films, WRRSM can reveal the crystal orientations in bulk samples, single crystals, highly strained or textured polycrystalline thin films (Inaba *et al.*, 2013; Nomoto *et al.*, 2016; Li *et al.*, 2022).

For polycrystalline thin films, Grazing Incident Diffraction (GID) is often employed to maximise the interaction between X-ray and the thin film layers. In order to work with a cradle tilting sample out of goniometer plane, ω-φ compensated Side Inclination Grazing Incident Diffraction (SI-GID) has been developed (Wang & van Riessen, 2017; Wang, 2021), which guarantees that the incident beam remains stationary in the sample coordinates. Essentially, ω-φ compensated SI-GID can cover 2θ scans in both out-of-plane direction, in-plane direction, as well as any directions in between (Figure 2 of Wang, 2021). If 2D frames are scanned in this geometry, WRRSM can be performed for thin film samples, as shown in Figure 2. Combining 2D frames from 2θ scans at χ = 0° (Figure 2a) and several χ steps (Figure 2b) allows the mapping of reciprocal space of thin film samples. This measurement is effectively equivalent to Grazing Incident Wide Range X-ray Scattering (GI-WAXS) measurement on Small Angle X-ray Scattering (SAXS) synchrotron beamline (Steele *et al.*, 2023) and some laboratory instruments (Kobayashi & Inaba, 2016) and has been applied to structure investigations of organic conductive thin film, exploring their molecular stacking orientations on substrates (Sethumadhavan *et al.*, 2024).

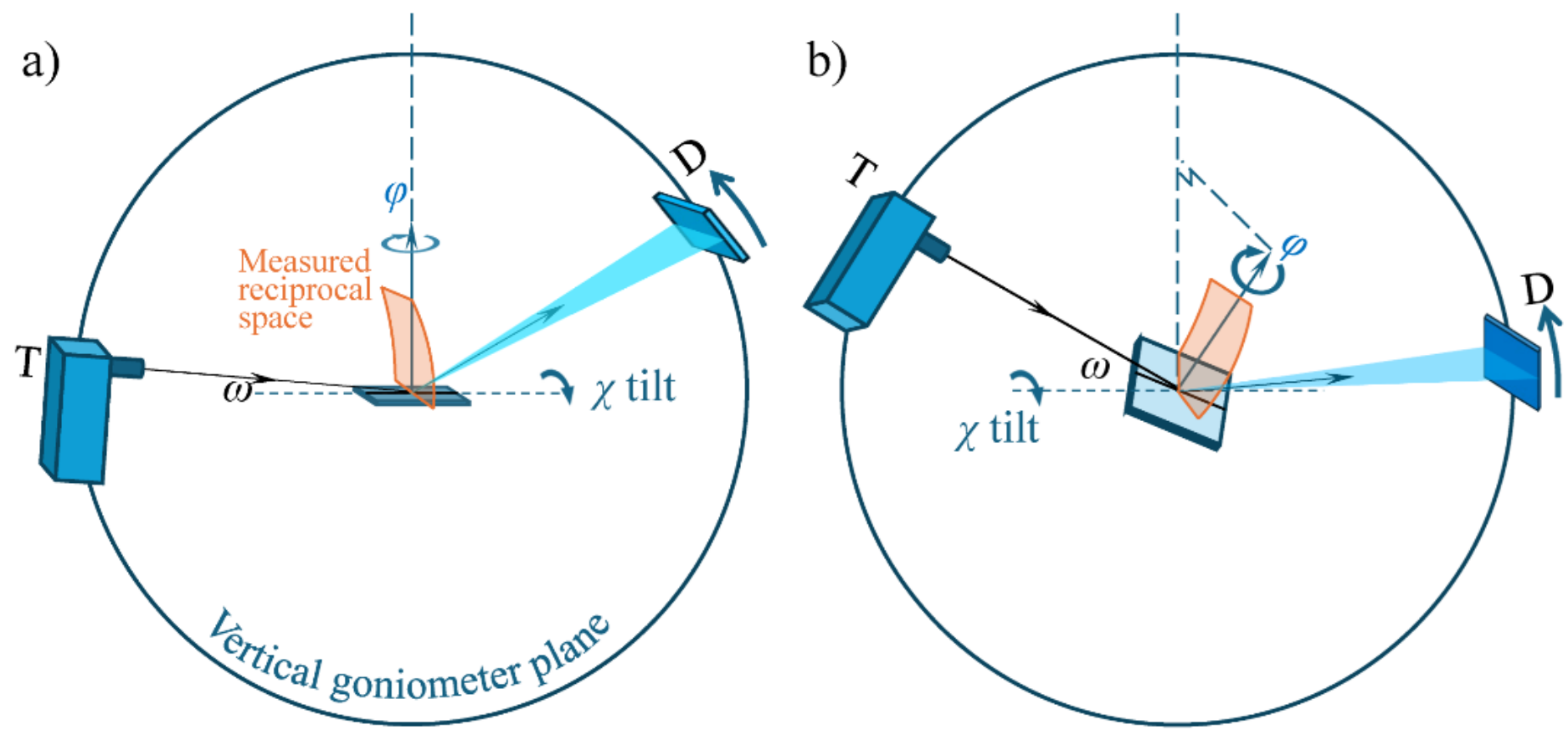

**Figure 2** WRRSM in ω-φ compensated SI-GID geometry: **a)** out-of-plane GID direction at χ = 0°; **b)** general GID direction at positive χ angles. T: X-ray Tube; D: Detector; Orange areas: the mapped reciprocal spaces at each χ tilt angle.

Existing commercial software for converting and merging 2D frames into WRRSM currently has several issue or limitations: 1) the calculated reciprocal space coordinates of the same *hkl* reflection on neighbouring frames seems not agree with each other (Supporting Document S1); 2) intensity reductions due to sample self-absorption are not corrected; 3) the output format of WRRSMs does not support follow-up quantitative analyses; 4) WRRSMs from opposite azimuthal directions cannot be combined into a whole upper semicircle of reciprocal space (Figure 1b); 5) the stitching edges of adjacent frames on the generated WRRSM are obvious, especially when applied on poorly crystalline or amorphous samples (Sethumadhavan *et al.*, 2024).

The following sections will report the geometric equations implemented in an open-source python program *RSstitcher* for generating WRRSM from 2D frames scanned in either symmetric mode or in ω-φ compensated SI-GID mode. The configurations of laboratory diffractometers and scan schemes, the algorithm and functionalities of *RSstitcher*, and the example WRRSMs generated from 2D frames collected using two common brands of laboratory X-ray diffractometers for each geometry are presented.

## 2. Geometries

In symmetric reflection geometry, also known as coupled θ/θ scan, the mapped reciprocal space plane is perpendicular to the sample surface (Figure 1b), while in ω-φ compensated SI-GID geometry, the merged WRRSM from 2D 2θ scans bends towards the X-ray tube (Figure 2). The angular relationships of both geometries are detailed separately as follows. These two 2D-scan geometries on laboratory XRD systems are fundamentally different to previous reported processing for static 2D frames collected from synchrotron beamlines (Baker *et al.*, 2010; Schrode *et al.*, 2019).

### 2.1. WRRSM in symmetric scans

As shown in the schematic geometry (Figure 3a), to measure the reciprocal plane OCC’ perpendicular to the sample surface, the X-ray tube “T” and a 2D detector are scanned symmetrically from low to high 2θ angle, recording a 2D diffraction frame in detector coordinates (x,y) (Figure 3b).

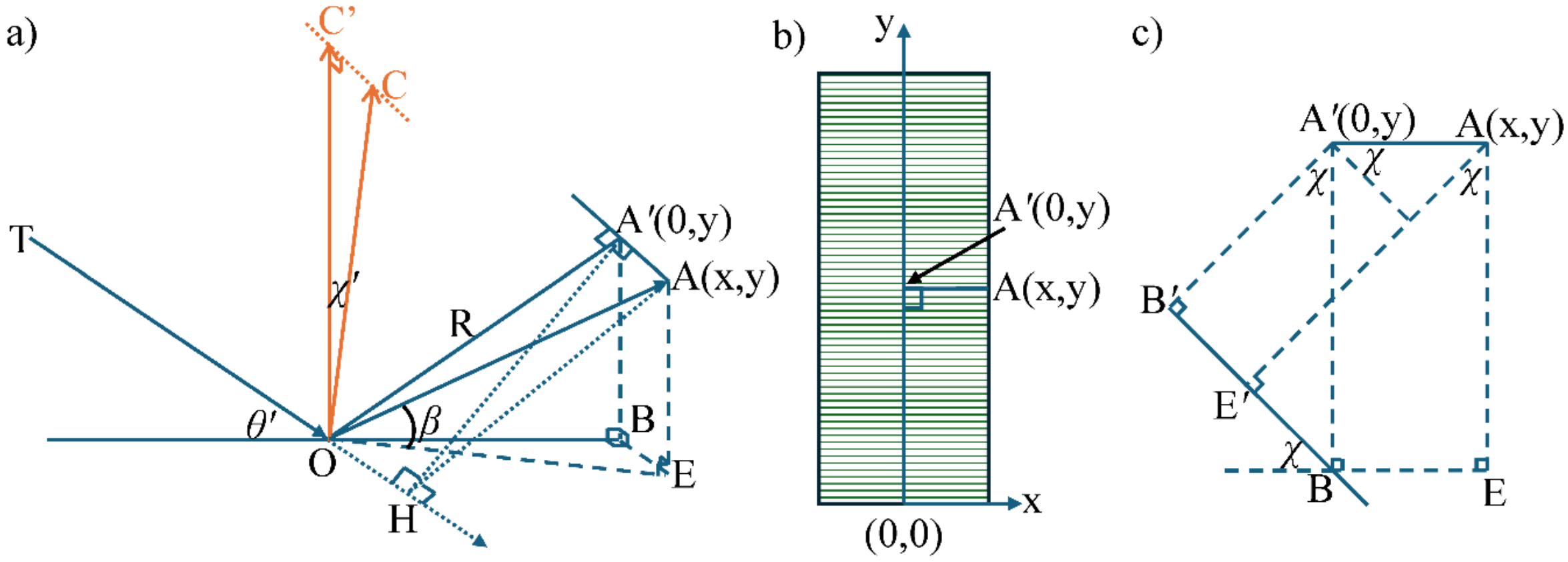

**Figure 3** **a)** Schematic drawing of 2D scans in symmetric geometry. The incident X-ray beam TOH and the diffracted X-ray beam OA’ lie in the goniometer plane, which intersects with the central line of 2D detector. The diffracted beam OA is out of the goniometer plane. The scattering vectors OC’ and OC in the reciprocal space (orange colour) corresponding to the diffraction beams OA’ and OA form a $\chi'$ angle, and they are drawn so that $|\overrightarrow{k_i}| = |\overrightarrow{k_f}| = R$, the goniometer radius. C’C is approximately perpendicular to the goniometer plane. AH and A’H are both perpendicular to the incident beam TH; A’B and AE are perpendicular to the horizontal sample surface at $\chi$=0°; $\beta$=∠AOE defines the take-off angle of diffracted beam from the sample surface; A’A, BE are perpendicular to the goniometer plane. **b)** In the recorded 2D frame, A’(0, $y$) lies in the detector centre line (x=0), while A($x$, $y$) denotes a general pixel, in which $x$ and $y$ are the pixel numbers in the collected 2D frame. **c)** Redrawn plane A’BEA from a) after sample cradle tilts $\chi$ angle around the OB axis, the perpendicular foots B and E shift to B’ and E’.

For the symmetric 2D scan geometry shown in Figure 3, the pixel numbers ($x$, $y$) of each pixel on the collected 2D frames can be converted into a coordinate ($S_x$, $S_z$) in reciprocal space, using below equations. For the pixels (0, $y$) on the centre line of the detector, the corresponding 2θ’ angles in radians are:

$$2\theta' = \angle A'OH = y * DCA \qquad (1)$$

, where $DCA$ stands for the Detector Channel Angle = $\tan^{-1}(\zeta/R)$, in which $\zeta$ is the pixel size of the detector. For each general pixel A(x, y) in the 2D frame, the corresponding 2θ angle in radians is:

$$2\theta = \angle\mathrm{AOH} = \cos^{-1}\left(\frac{|\mathrm{OH}|}{|\mathrm{OA}|}\right) = \cos^{-1}\left(\frac{R\cos 2\theta'}{\sqrt{R^2 + x^2\zeta^2}}\right) \quad (2)$$

, while the corresponding $\chi'$ angle in radians is:

$$\chi' = \angle\mathrm{C'OC} = \tan^{-1}\left(\frac{|\mathrm{C'C}|}{|\mathrm{C'O}|}\right) \approx \tan^{-1}\left(\frac{|\mathrm{A'A}|}{|\mathrm{C'O}|}\right) = \tan^{-1}\left(\frac{x\zeta}{2R\sin\theta'}\right) \quad (3)$$

As shown in Figure 1b, positive cradle tilt $\chi$ angles correspond to negative $\chi$ angle in the sample reciprocal space, therefore the zenithal angle $\psi$ between reciprocal space vectors and the sample normal corresponding to all combined frames can be calculated as:

$$\psi = \chi' - \chi \quad (4)$$

In the reciprocal space, the modulus of reciprocal space vector $|\mathrm{S}| = 1/d = 2\sin\theta/\lambda$, where $\lambda$ is the radiation wavelength. The horizontal components $S_x$ and the vertical component $S_z$ are therefore:

$$\mathrm{S_x} = |\mathrm{S}|\sin\psi\,;\ \mathrm{S_z} = |\mathrm{S}|\cos\psi \quad (5)$$

As shown in Figure 3c, after sample cradle tilts $\chi$ angle, the take-off angle of diffracted beam OA becomes $\beta$ =∠AOE', which can be derived from below equation:

$$\sin\beta = \frac{|\mathrm{AE'}|}{|\mathrm{OA}|} = \frac{|\mathrm{A'B}|\cos\chi + |\mathrm{A'A}|\sin\chi}{|\mathrm{OA}|} = \frac{R\sin\theta'\cos\chi + x\zeta\sin\chi}{\sqrt{R^2 + x^2\zeta^2}} \quad (6)$$

Eq.(6) can be calculated for every pixel in every 2D frame and assists to determine pixels below sample surface ($\beta$<0). The actual incident angle $\alpha$ between the incident beam and the sample surface changes with $\chi$ tilts, as shown in Eq.(7):

$$\sin\alpha = \sin\theta'\cos\chi \quad (7)$$

, which is derived from the previous report (Eq.(1) in Wang, 2021).

### 2.2. WRRSM in ω-φ compensated SI-GID mode

In the ω-φ compensated SI-GID geometry (Figure 2b), the incident beam is stationary relative to the sample regardless of χ tilts (Wang & van Riessen, 2017) and the incident angle is usually quite small. The 2D 2θ scans in this geometry are shown in Figure 4a, while the actual mapped reciprocal space is shown in Figure 4b.

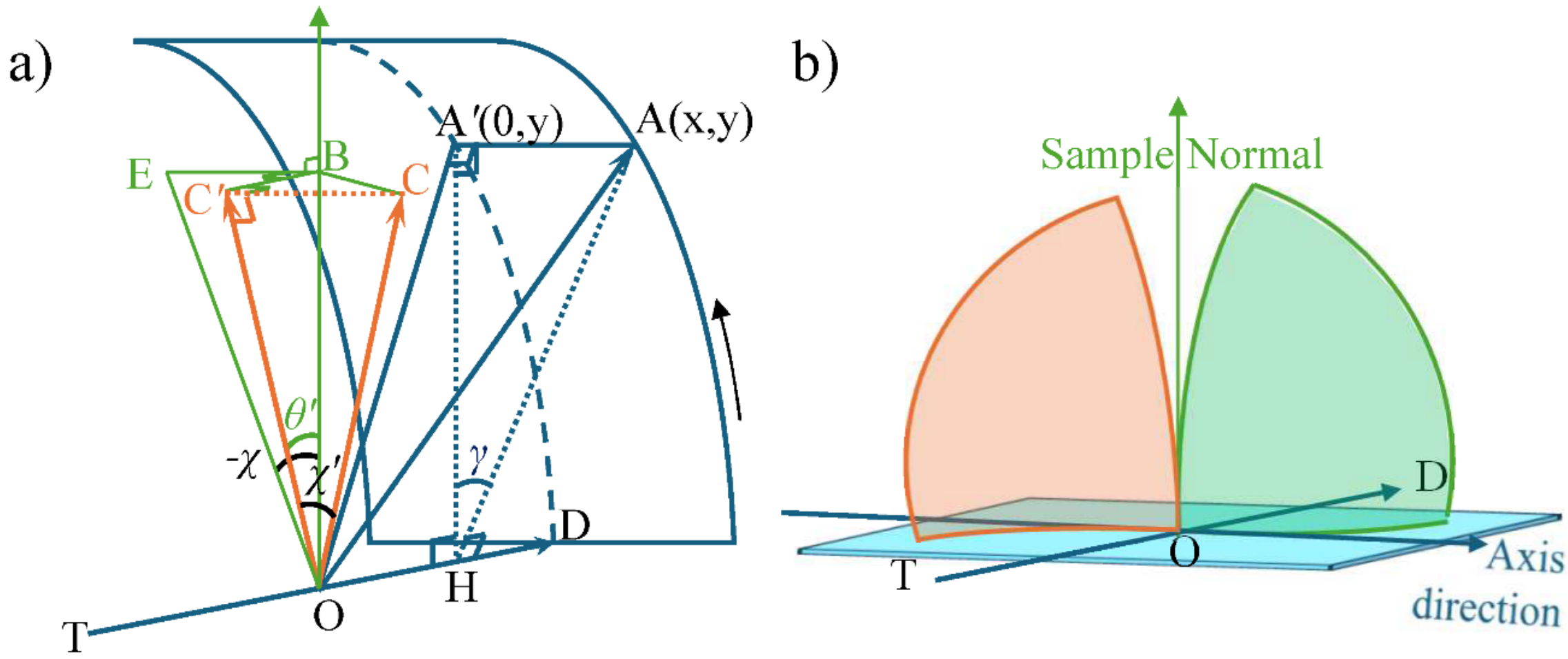

**Figure 4** **a)** Schematic geometry of 2D 2θ scans in ω-φ compensated SI-GID mode. A point focus beam from the X-ray tube T shoot a sample at the goniometer centre O, leaving the direct beam position D on the 2D frame. The scattering vectors OC' and OC in the reciprocal space (orange colour) corresponding to the diffraction beams OA' and OA form a $\chi'$ angle, and they are drawn so that $|\overrightarrow{k_i}| = |\overrightarrow{k_f}| = R$, the goniometer radius. CC' is approximately perpendicular to the goniometer plane; C'B is perpendicular to the sample normal OB before cradle tilt and sample normal OE after cradle tilt $\chi$ angle[1]; AH and A'H are both perpendicular to the direct X-ray beam TD; **b)** The mapped reciprocal space in the sample coordinate, the primary beam TOD is stationery. The orange side of reciprocal space is mapped in an azimuthal direction of $\varphi_0$, while the green side of reciprocal space is mapped in the opposite azimuthal direction of $\varphi = \varphi_0+180°$.

For the 2D 2θ scans in ω-φ compensated SI-GID mode shown in Figure 4a, the pixel numbers ($x$, $y$) of each pixel on the collected 2D frames can be converted to a coordinate ($S_r$, $S_z$), rather than ($S_x$, $S_z$) in reciprocal space using below equations. This is because the measured reciprocal space in GID is not in any plane, but a surface gradually bends towards the X-ray source, as shown in Figure 4b. Practically, an intensity map of normal direction $S_z$ and radial direction $S_r$ is mapped like GI-WAXS frames. A missing wedge will be shown in this $S_z$-$S_r$ map, since the reciprocal space in sample normal direction cannot be covered by any grazing incident measurement. For the pixels (0, $y$) on the centre line of the detector, the calculation of the corresponding 2θ' angles in radians are same as Eq.(1). For each general pixel A($x$, $y$) in the 2D frame, the corresponding 2θ angle in radians also follows Eq.(2). The corresponding $\chi'$ angles $\angle C'OC$ in radians also follow the Eq.(3). The cradle χ tilt axis is very close to the X-ray beam direction TOD in Figure 4a, therefore the χ angle is in the same plane of γ angle $\angle \mathrm{A'HA}$ which can be calculated as Eq.(8):

$$\gamma = \angle \mathrm{A'HA} = \tan^{-1}\left(\frac{x\zeta}{R\sin 2\theta'}\right) \tag{8}$$

[1] When the cradle tile $\chi$ angle is positive, sample normal OE should be tilted to the right side of OB in Figure 4a). For clarity, Figure 4a) shows the case when $\chi$ angle is negative, which leads to sample normal OE be tilted to the left side of OB, with "-$\chi$" angle marked.

The zenithal angle in sample reciprocal space ($\angle$EOC in Figure 4a) can be calculated as Eq.(9) (detailed in Appendix A):

$$\angle \mathrm{EOC} = \cos^{-1}(\cos\chi\cos\theta'\cos\chi' + \sin\chi\sin\chi') \tag{9}$$

The sample surface radial components $S_r$ and the sample normal component $S_z$ in reciprocal space are:

$$\mathrm{S_r} = |\mathrm{S}|\sin\angle\mathrm{EOC};\ \mathrm{S}_z = |\mathrm{S}|\cos\angle\mathrm{EOC} \tag{10}$$

Since the incident angle ω in GID is usually small, the take-off angle of diffracted beam recorded by each pixel in the 2D scans collected in ω-φ compensated SI-GID mode can be approximately calculated as Eq.(11):

$$\sin\beta = \frac{R\sin(2\theta' - \alpha)\cos\chi + x\zeta\sin\chi}{\sqrt{R^2 + x^2\zeta^2}} \tag{11}$$

, in which $\alpha$ can be calculated from Eq.(12):

$$\sin\alpha = \sin\omega\cos\chi \tag{12}$$

The $\alpha$ angle is the actual incident angle kept constant in the ω-φ compensated SI-GID geometry, which needs only be calculated once. Eq.(11) is similar to Eq.(6), except the take-off angle of the diffraction beam OA' at χ = 0° become $2\theta' - \alpha$.

### 2.3. Intensity correction due to sample self-absorption

It can be imaged from Figure 1b that the diffraction beams on the Debye arc recorded by 2D detectors have gone through different beam paths inside the sample. The lower the take-off angles of the diffraction beam, the higher the sample self-absorption of the diffraction intensity. The same effect also happens in the 2D scans in the ω-φ compensated SI-GID geometry (Figure 2b). Therefore, the diffraction intensities recorded in each 2D scanned frames require an intensity correction. The primary and secondary beam paths inside a flat sample depend only on the actual incident angle $\alpha$ of the primary beam and the actual take-off angle $\beta$ of the diffract beam, respectively. A reverse use of the Eq.(4) reported in Rowles & Buckley (2017) can be applied as Eq.(13):

$$I_{cor} = \frac{I}{2}\left(1 + \frac{\sin\alpha}{\sin\beta}\right) \tag{13}$$

, in which $I_{cor}$ is the intensity after correction, while $I$ is the raw intensity recorded by each pixel. The terms "$\sin\beta$" and "$\sin\alpha$" have been derived in Eq.(6,7) respectively for symmetric geometry, and in Eq.(11,12) respectively for the ω-φ compensated SI-GID geometry. The effect of this intensity correction will be discussed in Section 6.1.

### 3. Samples and Frame collection schemes

The previously reported fibre textured thin-section of satin spar sample (Wang et al., 2024) was used to demonstrate the WRRSM generation from 2D frames collected in symmetric scans. The parallel feature of reciprocal space lines of this fibre textured bulk sample assists a confirmation of correct WRRSM generation. The previously reported 100 nm thick PEDOT, *i.e.* Poly(3,4-ethylenedioxythiophene), layer on PET substrate (Sethumadhavan et al., 2024) was used to demonstrate WRRSM generation from 2D frames collected in ω-φ compensated SI-GID geometry. A flattened NIST SRM676a corundum powder surface and a NIST SRM1976a corundum ceramic disk were measured in symmetric geometry to demonstrate the intensity integration functions of *RSstitcher* software.

#### 3.1. Optimisation of χ tilt angles

A key scan scheme parameter needs be calculated for WRRSM data collection is the χ step size of cradle tilt angle. An optimised χ step size should leave no χ gap between each 2D frames at the upmost 2θ angle (the outer rim of the mapped reciprocal space in Figure 1b), meanwhile minimise the number of 2D frames required to cover the reciprocal space plane. According to Eq. (3), the $\chi'$ angle of the corner pixel furthermost to the direct beam origin on each 2D frame can be calculated according to the detector width in the goniometer axis direction and the highest 2θ angle:

$$\text{For Hypix3000 detector: } \chi'_{\text{corner}} = \tan^{-1}\left(\frac{77.5/2\ mm}{2*121\ mm \sin(140°/2)}\right) \approx 9.76°;$$
$$\text{For LynxEye XE-T detector: } \chi'_{\text{corner}} = \tan^{-1}\left(\frac{14.4/2\ mm}{2*280\ mm \sin(125°/2)}\right) \approx 0.83° \quad (14)$$

, in which the half widths of the detectors (77.5/2 mm and 14.4/2 mm) are substituting $x\zeta$ in Eq. (3); the secondary goniometer distance in Table 1 (121 mm and 280 mm) are substituting $R$ in Eq. (3); half of the stop 2θ angle in Table 1 (140°/2 and 125°/2) are substituting $\theta'$ in Eq. (3). A cradle tilt $\chi$ step size for each detection system can then be planned to be just below the corresponding $2\chi'_{\text{corner}}$ values, as used in the "Scan Scheme" rows of Table 1.

Analogically, the $\gamma$ angle of the corner pixel furthermost to the direct beam origin of each 2D frame can be calculated using Eq.(8) for GID scans:

$$\text{For Hypix3000 detector: } \gamma_{\text{corner}} = \tan^{-1}\left(\frac{77.5/2\ mm}{121\ mm \sin 40°}\right) \approx 26.48°$$
$$\text{For LynxEye XE-T detector: } \gamma_{\text{corner}} = \tan^{-1}\left(\frac{14.4/2\ mm}{280\ mm \sin 45°}\right) \approx 2.08° \quad (15)$$

A cradle tilt $\chi$ step size for each detection system can then be planned to be just below the corresponding $2\gamma_{\text{corner}}$ values, as used in the "Scan Scheme" rows of Table 2.

#### 3.2. Scan schemes

Two brands of common laboratory X-ray diffraction systems were used to collect 2D frames which were combined into WRRSMs using the developed *RSstitcher* program. The configurations used to collect WRRSM in symmetric 2D scans are summarised in Table 1.

**Table 1** The configurations of the X-ray diffraction systems used for WRRSM data collection in symmetric scans.

| Diffractometers | Rigaku SmartLab | Bruker D8 Advance |
| --- | --- | --- |
| X-ray sources | Cu sealed tube (line focus, 40kV 40mA) | Co twist tube (point focus, 35kV 40mA) |
| Beam Conditionings | CBO-PB0.5mm, CBO-f, 0.8 mm collimator | Poly-capillary module, 0.5 mm collimator |
| Goniometer radius | Primary: 300 mm<br>Secondary: 121 mm | Primary: 280 mm<br>Secondary: 280 mm |
| Sample stages | χ-φ Cradle on standard motorised Z stage | Compact χ-φ-z Cradle Plus |
| Detectors | Hypix3000 (Fluorescence suppression mode, 2D scan);<br>Pixel size: 100 μm;<br>W: 775 pixels;<br>H: 385 pixels | LynxEye XE-T (90° rotated, High Energy Resolution mode, 2D scan);<br>Pixel size: 75 μm;<br>W: 192 pixels;<br>H: 1.2 mm detector slit |
| Scan schemes | Symmetric 2D scans: 5 – 140 °2θ;<br>χ: from 9 to 81° at 18° step size;<br>φ: at 0° and 180°;<br>In total 10 frames;<br>Each frame was collected in 7 mins. | Symmetric 2D scans: 15 – 125 °2θ, Step size: 0.1°;<br>χ: from 0.8 to 79.2° at 1.6° step size;<br>φ: at 0° and 180°;<br>In total 100 frames;<br>Each frame was collected in 18 mins. |

Similarly, the configurations used to collect WRRSM in the ω-φ compensated SI-GID mode are summarised in Table 2. The ω and φ compensation angles at each χ tilt are calculated using the Eq. (1) and Eq. (2) reported previously (Wang, 2021). Under these scan schemes, the maximum φ compensation angel is less than 5°:

$$\varphi_{max} = \sin^{-1}(\tan\chi_{max} * \tan\alpha) = \sin^{-1}(\tan 77.9° \times \tan 1°) \approx 4.71° \qquad (16)$$

, in which $\chi_{max}$ is the maximum cradle tile angle, $\alpha$ is the effective incident angle.

**Table 2** The configurations of the X-ray diffraction systems used for WRRSM data collection in ω-φ compensated SI-GID mode.

| Diffractometers | Rigaku SmartLab | Bruker D8 Advance |
| --- | --- | --- |
| X-ray sources | Cu sealed tube (line focus, 40kV 40mA) | Co twist tube (point focus, 35kV 40mA) |
| Beam Conditionings | CBO-PB0.5mm, CBO-f, 0.3 mm collimator | Poly-capillary module, 0.5 mm collimator |

| Goniometer radius | Primary: 300 mm<br>Secondary: 121 mm | Primary: 280 mm<br>Secondary: 280 mm |
|---|---|---|
| Sample stages | χ-φ Cradle on standard motorised Z stage | Compact χ-φ-z Cradle Plus |
| Detectors | Hypix3000 (Fluorescence suppression mode, 2D scan);<br>Pixel size: 100 μm;<br>W: 775 pixels;<br>H: 385 pixels | LynxEye XE-T (90° rotated, High Count Rate mode with Kβ filter, 2D scan);<br>Pixel size: 75 μm;<br>W: 192 pixels;<br>H: 1.2 mm detector slit |
| Scan schemes | ω-φ compensated maintaining an incident angle $\alpha$ = 0.4° in a same sample azimuthal direction;<br>2D 2θ scans: 5 – 40 °2θ;<br>χ: 25 to 75° at 50° step size;<br>$\varphi_0$: at 0° and 180°;<br>In total 4 frames;<br>Each frame was collected in 37 mins. | ω-φ compensated maintaining an incident angle $\alpha$ = 1° in a same sample azimuthal direction;<br>2D 2θ scans: 3 – 45 °2θ; Step size: 0.1°;<br>χ: 1.9 to 77.9° at 3.8° step size;<br>$\varphi_0$: at 0° and 180°;<br>In total 42 frames;<br>Each frame was collected in 21 mins. |

## 4. Algorithm and functionalities

The algorithm was implemented in python language with *NumPy* (Harris et al., 2020), *SciPy* (Virtanen *et al.*, 2020), *pandas* (The pandas development team, 2025) and *tifffile* (Gohlke, 2024) libraries. In the initial release, 2D frames in either .img or .gfrm format are imported with the *FabIO* python library (Knudsen *et al.*, 2013) from a single parent directory for each scan. The program automatically loads all 2D frames in all subdirectories. Experimental information including sample to detector distance, wavelength, detector channel angle (DCA), detector pixel size, and the omega, phi, and chi angles are extracted from the file header. The initial phi angle ($\varphi_0$) is fixed to the first file, with the positive $S_x$-axis parallel to the azimuthal direction of $\varphi_0$.

The reciprocal space coordinate transformations are performed using vectorised operations. For visualising ω-φ compensated SI-GID data, the intensities at the edge of each frame can be optionally blurred with a 2D Gaussian filter of a sigma radius equal to 10% (default value, customisable) of the short edge of each frame, to reduce the artifacts of the overlapping area of adjacent frames. The coordinates of each detector pixel are converted to reciprocal space ($S_x$, $S_z$) according to Eq.(5). However, to ensure the 2D frames measured at $\varphi = \varphi_0+180°$ are correctly merged with 2D frames measured at $\varphi_0$, a multiplier (M) is applied to $S_x$, which equals to one when $\varphi = \varphi_0$ and equals to negative one when $\varphi = \varphi_0 + 180$, within a tolerance of $\varphi_{max}$ in Eq.(16) (default value 5°, customisable) to allow for ω-φ compensation in SI-GID geometry:

$$S_x = M|S|\sin\psi \tag{17}$$

, in which M reverses the sign of $S_x$ for 2D frames measured at $\varphi = \varphi_0+180°$.

For symmetric 2D frames, the intensity data of each reciprocal space coordinate is then combined by taking the average intensity value of the corresponding pixels from all individual 2D frames that correspond to the reciprocal space pixel. For ω-φ compensated SI-GID frames the maximum value is taken to reduce edge effects where edge blurring is implemented. The reciprocal space pixel size (ΔS) is defined according to the theoretical resolution limit of the experiment:

$$\Delta S = \frac{2sin(DCA)}{\lambda} \quad (18)$$

, which is then rounded to one significant figure. Data from a single raw data pixel is placed in the closest reciprocal space pixel without pixel splitting. While pixel splitting would in theory result in higher fidelity results, in practice we find pixel splitting an unnecessary computational overhead.

The intensity data either in a linear (default), logarithmic, or square root scale is then exported as a quantitative 32-bit tiff image. Also exported is a separate tiff image file of the same pixel size with the location of the reciprocal space origin, the $S_x$ and $S_z$ axes, together with rings at a default interval of $\Delta S = 0.1$ Å$^{-1}$, to facilitate user's interpretation and further annotations in other tiff image process software *e.g.* ImageJ (Schindelin *et al.*, 2012).

Other than exporting WRRSM for visual appreciation and qualitative evaluations, the program is also able to assist quantitative analysis of the profile and deformation of combined Debye arcs for symmetric frames. These functions include $S$ integrations of combined 2D reciprocal space map over a range of zenithal directions, or $\chi_S$ integration of a particular Debye arcs, respectively. The former allows exportation of 1D data of diffraction intensities on S (= 1/d) scale for further XRD analysis including phase identification and quantitative phase analysis, while the latter integration exports the intensity profile along Debye arcs, based on user's input of a set of S ranges of the target Debye arcs. Crystal texture, spottiness due to coarse grains, and epitaxial layer qualities or mosaicities can be studied from the exported Debye arc profiles.

## 5. Software distributions and operations

There are two distributions the *RSstitcher* program maintained: A python open-source program and a webpage-based tool free from installation. Both distributions share similar functions, with the former supporting further development and the latter focused on ease of use and visualisation.

### 5.1. Python program distribution

The software *RSstitcher* is available in the **Data Availability** Section. The program can be installed with *pip* command in Python v3.13. It has also been packaged as single binary executables using *PyInstaller* (Cortesi *et al.*, 2025) for computer operation systems of Windows, macOS, and Linux. It runs in command-line interface (CLI) and typically uses up to 1 GB of RAM.

In CLI, users must provide a path to a directory containing all the 2D diffraction frames in either Bruker (.gfrm) or Rigaku (.img) format with additional formats that are supported by *FabIO* available through user defined configuration files. The user may then specify result filenames. For example, below command is used to generate the RS-RSM for the example 2D frames in .gfrm format:

```
C:\RSstitcher>RSstitcher tests/data/bruker_symmetric --write
pixels_tiff=output.tiff --write grid_tiff=grid.tiff
```

, which prompts the software to print out following experiment parameters together with the outputs:

```
Experiment parameters:
----------------------
Type                          gfrm
Data size                     192 x 1101 pixels
Detector distance             280.0 mm
Phi 0                         90.0 degrees
Wavelength                    1.78897 Å
Pixel size                    0.075 mm
Theta pixel                   0.100 degrees
Phi tolerance                 5.0 degrees
Blur                          19 pixels
Delta s                       0.002 Å⁻¹
Rounding                      3 decimal places
Scaling                       linear

Results:
--------
Sx range                      -0.962 to 0.972 Å⁻¹
Sz range                      -0.004 to 0.990 Å⁻¹
Number of pixels              968 x 498 pixels
Number of images processed    100 images
Time taken                    1.08 seconds
```

The user can then view the generated WRRSM in 32-bit tiff file under the user specified name of “output.tiff”, together with a generated “grid.tiff” file of the same pixel dimensions containing the reciprocal space grid information.

A list of Options and Helps are available by running “`C:\RSstitcher>RSstitcher --help`”.

### 5.2. Webtool of *RSstitcher*

The webpage “https://eresearchqut.github.io/*RSstitcher*/” contains the latest version of *RSstitcher* and runs entirely on users’ local computers (Figure 5). Raw data and results remain entirely on the local computer and are not transferred to an external server.

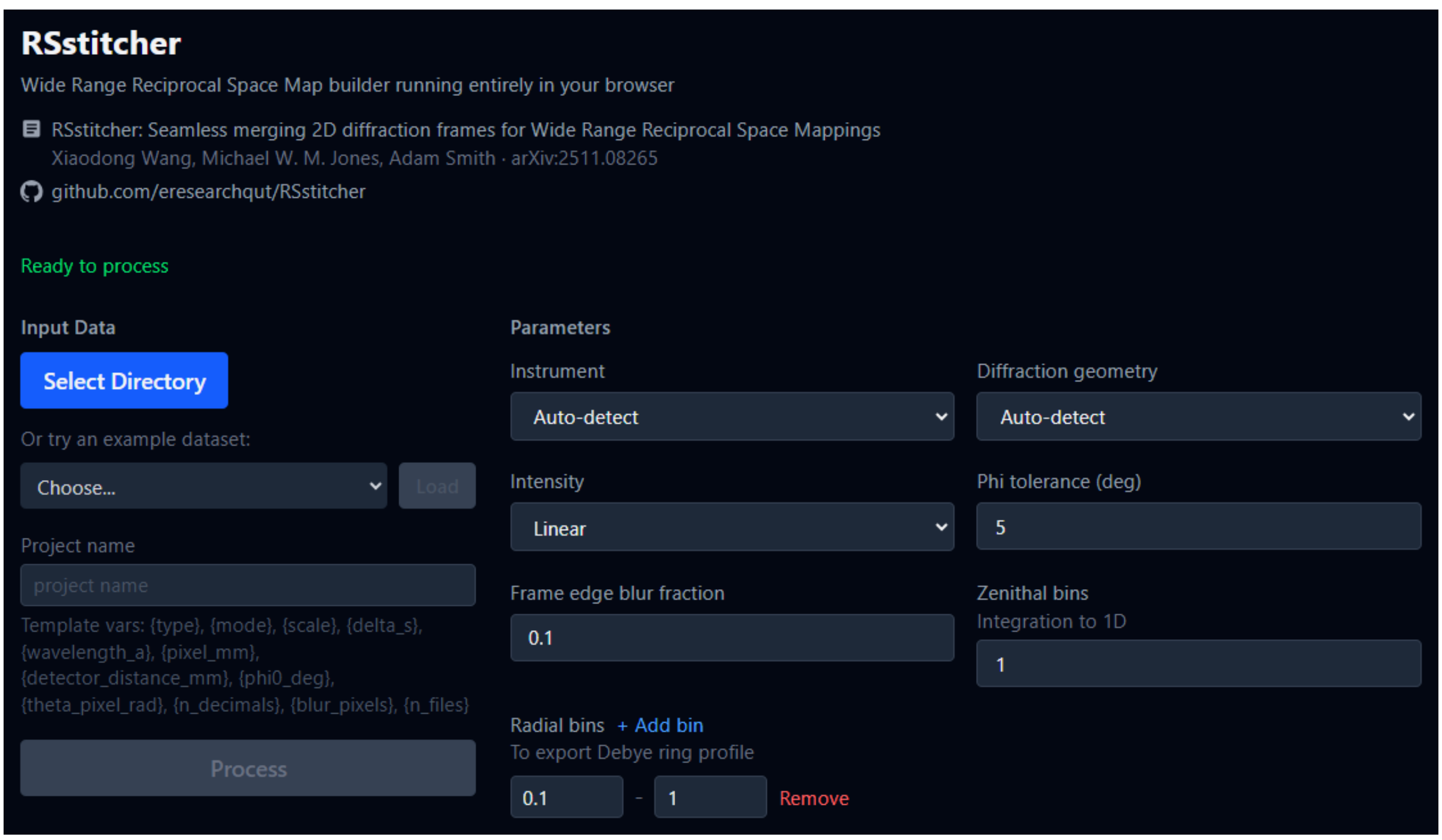

**Figure 5** The Graphic User Interface (GUI) of the webpage based *RSstitcher*. A full GUI including the result sections can be found in the Supporting Document S2.

Users can choose example dataset from the dropdown list, or able to select the file path of a folder containing raw 2D frame data in either .img or .gfrm formats from local disks. For other 2D frame formats supported by the *FabIO* python library, users can write an instrument config .json file according to the instruction pops up when “Custom” is chosen in the “Instrument” dropdown list. Users only need to click the “Process” button and the combined WRRSM is displayed in the same webpage. The *RSstitcher* program is able to automatically detect whether the series of 2D frames was collected from symmetric 2D scans or from ω-φ compensated SI-GID measurements, by checking whether the ω values in the whole series of 2D frames is a constant in the former or has changed for compensation in the latter measurements. Users can force the program to use the corresponding algorithms in Section 2.1 or 2.2 by choosing “Symmetric” or “GID” from the dropdown list. Three options of intensity scale: Linear, Square root, and Logarithmic are available in case more features need been shown for highly crystalline sample with strong reflections. Changing of “Phi tolerance” is only needed if the maximum φ compensation for ω-φ compensated SI-GID measurements is calculated to be over 5 deg in Eq.(16). The default “Frame edge blur fraction” value of 0.1 (between 0 and 1) turns out to work well for most of samples but can be increased if the overlapping edges of 2D frames are too obvious for some amorphous or semicrystalline samples.

The integration function to 1D data is executed for symmetric scan frames according to the integer number provided to the zenithal bin. Normally “1” zenithal bin is used to integrate all the available data into a single 1D XRD pattern. When sample with residual stress is measured, higher numbers of zenithal bins can be chosen to compare the S values of the same *hkl* peaks at different zenithal angles. The integration in $\chi_s$ direction for Debye arc profile can be activated by clicking “Add bin” and provide the lower and upper S values of the target Debye arc. The profile of multiple Debye arcs can be integrated simultaneously.

## 6. Results

The WRRSM images generated by *RSstitcher* in tiff format were then merged with the reciprocal space S grids in ImageJ software (Schindelin *et al.*, 2012). while the tiff images of S unit grids generated by *RSstitcher* were annotated with axes name and tick unit values. Since both tiff images are of same pixel dimensions, they can be readily merged into different colour channels using ImageJ software. The final WRRSMs presented in this section are composite images merged from the annotated WRRSM tiff images (in red channel) and the annotated S grid tiff images (in gray channel).

### 6.1. Effectiveness of intensity correction for sample self-absorption

The effectiveness of this intensity correction can be verified using the WRRSM of a corundum powder sample (NIST SRM 676a), which is supposed to provide constant intensity along Debye arcs. As shown in Figure 6a, the uncorrected intensities along the Debye arcs tend to be weaker as $\chi_s$ goes to higher angle. The absorption of the integrated 104 peak intensity (Figure 6c) becomes apparent when $|\chi_s|$ is higher than 40°. The corrected WRRSM in Figure 6b does show more even intensity along all the Debye arcs. The 104 peak intensities in Figure 6d were integrated from the corrected WRRSM return to constant. A full WRRSM of NIST SRM 676a is shown in Supporting Document S3.

It should be noted that at extreme low take-off angle $\beta$, the absorbed intensity recorded by 2D scans can reduce to the noise level. The intensity correction devised in Eq.(13) will not be able to recover those intensities. Therefore, intensities recorded by pixels below a $\beta$ cut-off angle on each 2D frames were not used in WRRSM generation. It can be imaged that the choice of $\beta$ cut-off value depends on the roughness of sample surface, therefore allowed to be customised by users in *RSstitcher*. A $\beta$ cut-off angle of 1.5° was applied to the pressed surface of corundum powder (NIST SRM 676a) for the WRRSMs shown in Figure 6. At low $\beta$ angles the intensity correction can still be significant, leading to the amplification of background noise, as can be seen at the bottom edge of Figure 6b.

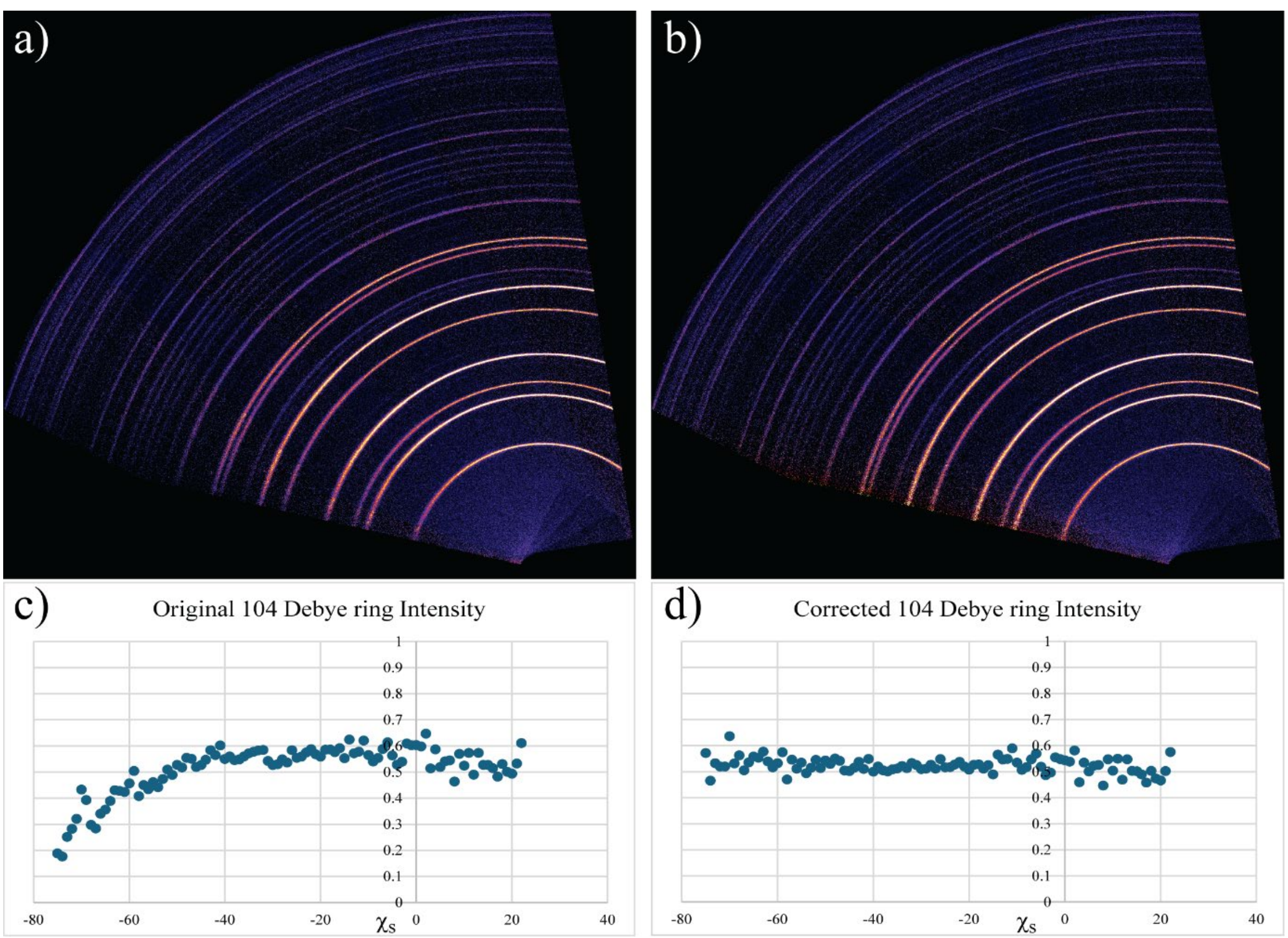

**Figure 6** Comparison of WRRSMs of corundum powder (NIST SRM 676a) **a)** without and **b)** with intensity correction of Eq.(13). The integrated intensities along the 104 Debye arcs are compared **c)** before and **d)** after this intensity correction.

#### 6.2. WRRSM from Symmetric 2D scans of fibre textured satin spar cross-sections

Figure 7a and 7b show similar WRRSM results collected on different XRD systems (Table 1) for the same cross-section sample of satin spar. The parallel reciprocal space “lines” formed by spots of Miller indexes *nkl* (*n* = 1, 2, 3, 4, 5, 6, 7) indicates the microstructure of fibre texture in this satin spar sample, which has been explained previously (Wang et al., 2024). A slight tilt of approximate 2° of these parallel reciprocal space “lines” in Figure 6a was from the miscut of this sample from the original satin spar rod. This example demonstrates the benefit of being able to merge WRRSMs at both $\varphi_0$ and $\varphi_0$ + 180° azimuthal directions, since they are not always in mirror symmetry. The measurable S region in Figure 7a is wider than that in Figure 7b, because a longer X-ray wavelength was used in the latter measurement.

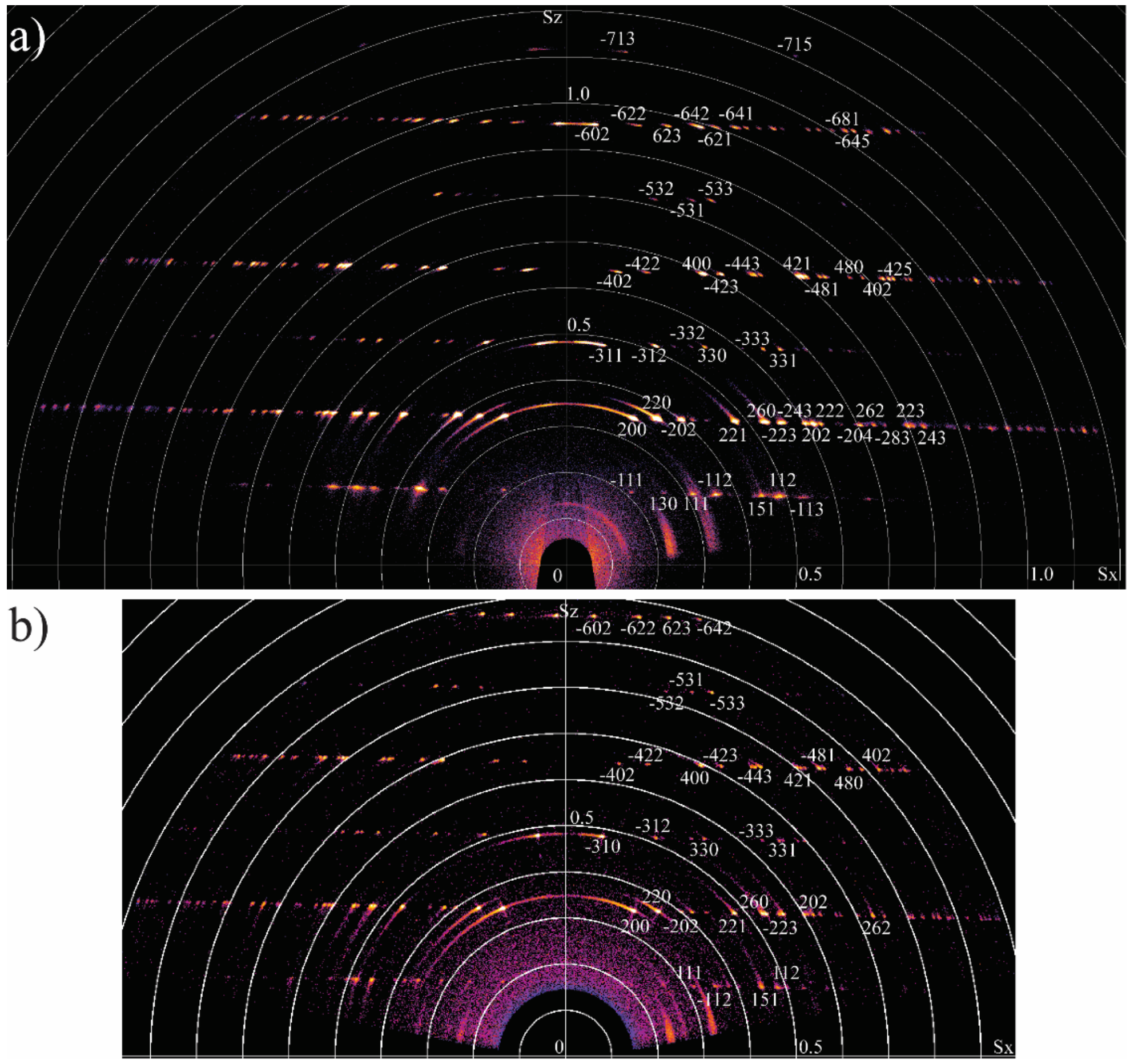

**Figure 7** WRRSMs of the cross-section of fibre textured satin spar, merged from symmetric 2D frames collected at different cradle $\chi$ tilts using a) the Hypix3000 detector setup and CuKα radiation in Table 1; b) the LynxEye XE-T detector setup and CoKα radiation in Table 1. The intensities are both shown in linear intensity scale. The reciprocal spaces of a) and b) are shown in the same S scale with the spacing between circular ticks of $\Delta s = 0.1$ Å$^{-1}$. The Miller indexes follow the gypsum unit cell definition in ICDD PDF No. 04-015-8262 (Comodi et al., 2008).

### 6.3. WRRSM from 2θ 2D scans in ω-φ compensated SI-GID geometry

Similar WRRSMs of the PEDOT thin film sample on PET substrate measured on the two XRD systems (Table 2) are shown in Figure 8. As in GI-WAXS plots, there are wedge-shaped regions along $S_z$ not measurable in the $S_z$ - $S_r$ plots of WRRSM results. The missing wedge in Figure 8b is wider than that in Figure 8a, because Figure 8b was collected using a longer X-ray wavelength (CoKα), which corresponds to a smaller Ewald sphere than that in Figure 8a. This polymer PET is poorly crystalline, though still possible to index the diffraction peaks of the PET substrate and the thin PEDOT layer with their corresponding Miller indices. This type of thin film characterisation is similar to what GI-WAXS (Grazing Incident Wide Angle X-ray Scattering) presents, which is typically only accessible on SAXS/WAXS instruments or synchrotron beamlines. With the geometry and python code developed in this report, the reciprocal space information of thin film samples can be mapped using any laboratory X-ray diffractometer equipped with a goniometer cradle and a 2D detector. The residual radial stripes in Figure 8b are the result of an increased density of non-zero values in regions where the frames overlap after they have been blurred with a 2D Gaussian filter of a sigma radius of 10% (default, customizable) of the width of each frame. The larger Hypix3000 detector used in Figure 8a results in a more gradual blur resulting in less obvious overlapping.

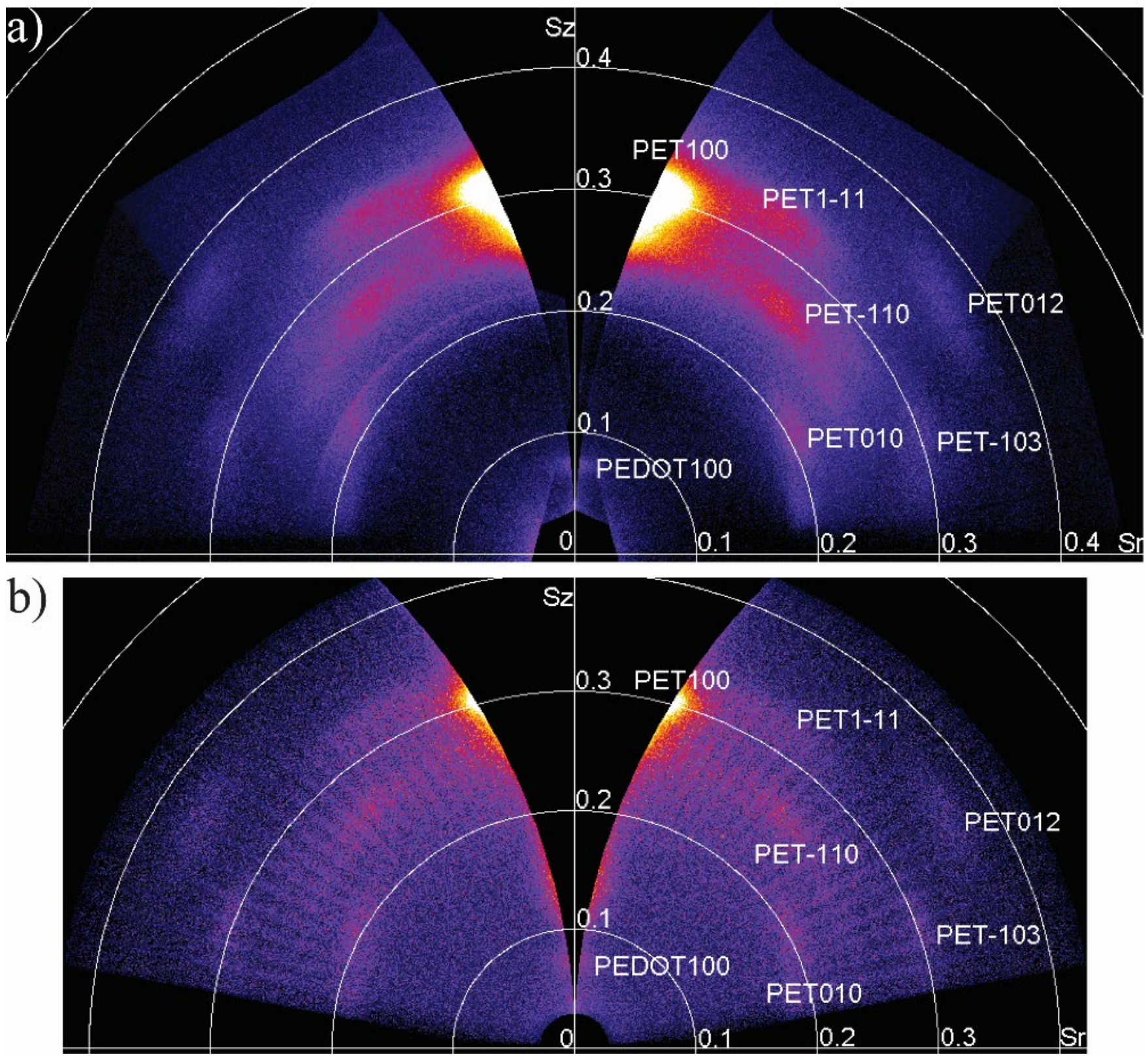

**Figure 8** The generated WRRSMs of the PEDOT on PET sample collected in ω-φ compensated SI-GID mode using a) Hypix3000 detector setup and CuKα radiation in Table 2; b) LynxEye XE-T detector setup and CoKα radiation in Table 2. The intensities are both shown in linear intensity scale. The reciprocal spaces of a) and b) are shown in the same S scale with the spacing between circular ticks of $\Delta s = 0.1$ Å$^{-1}$. The PET Miller indexes follow the unit cell definition in ICDD PDF No. 02-072-2446 (Daubeny et al., 1954).

One of the pitfalls of collecting 2D scans in GID geometry is the defocusing effect of the long footprint of X-ray beam on the sample surface. Due to the nature of grazing incidence of the primary point focus beam, the diffraction location is not only at the goniometer centre but also on a line stretching in the beam direction, turning the recorded Debye arcs into Debye belts. This geometric defocusing effect is most severe at 90 °2θ, while less obvious for low angle reflections (Rowles & Madsen, 2010). Avoiding extremely small incident angle or large beam collimator could help shortening the X-ray footprint length. And installing a large radial Soller slit in front of the 2D detector at a longer sample to detector distance could also help screening out reflections not from goniometer centre. In any doubt, running a standard material with similar peak positions in the same geometry is always a good practise and recommended.

### 6.4. Integration functions of *RSstiticher*

The WRRSM measured for a NIST SRM 1976c corundum ceramic disk is shown in Figure 9a. The reciprocal space circular ticks are omitted for clarity. The strong and weak intensity part of the same Debye arcs can be clearly visualised. Figure 9b shows the 1D patten integrated from the WRRSM aligning with the ICDD PDF card 04-004-2852 corundum phase. It can be seen that the measured relative intensities of different *hkl* reflections do not matches well with the relative diffraction intensities of powder diffraction pattern in the PDF card, indicating crystal texture of the NIST SRM 1976c corundum ceramic disk.

#### 6.4.1. Intensity profile along Debye arcs and Pole Figures

The integrated intensity profiles of some main Debye arcs are shown in Supporting Document S4. Due to the known axisymmetric texture nature of NIST SRM 1976c corundum ceramic disk (National Institute of Standards and Technology, 2019), pole figures of this SRM can be created by rotating the intensity profiles of Debye arcs 180° along the sample normal. Figure 10 explains how pole figures are synthesized from WRRSMs. In conventional texture measurements (Figure 10a), the intensity on selected reciprocal space shells (pole figures) were scanned. Since the cradle $\chi$, $\varphi$ drives are slow in speed and need to step into prescribed sample orientations for a 1D scan across a *hkl* peak profile, conventional texture measurement often takes dozens of hours. However, as shown in Figure 10b), the same upper hemispheric reciprocal space can be measured by a series of WRRSMs in different azimuthal directions. If a sample texture is in axis-symmetry, for example a spin-coated thin film and the NIST SRM 1976c corundum ceramic disk, *i.e.* texture is azimuthally equivalent, just one WRRSM is needed to create Pole Figures for many *hkl* reflections.

The pole figures of 012, 104, 110, 006, 113 and 116 reflections derived from the WRRSM in Figure 9a are shown in Figure 9c-9h, respectively. These pole figures match with previous reported pole figure for NIST SRM 1976a (Borovin *et al.*, 2018) and those of corundum ceramic plate synthesized in similar procedure (Medendorp *et al.*, 1995).

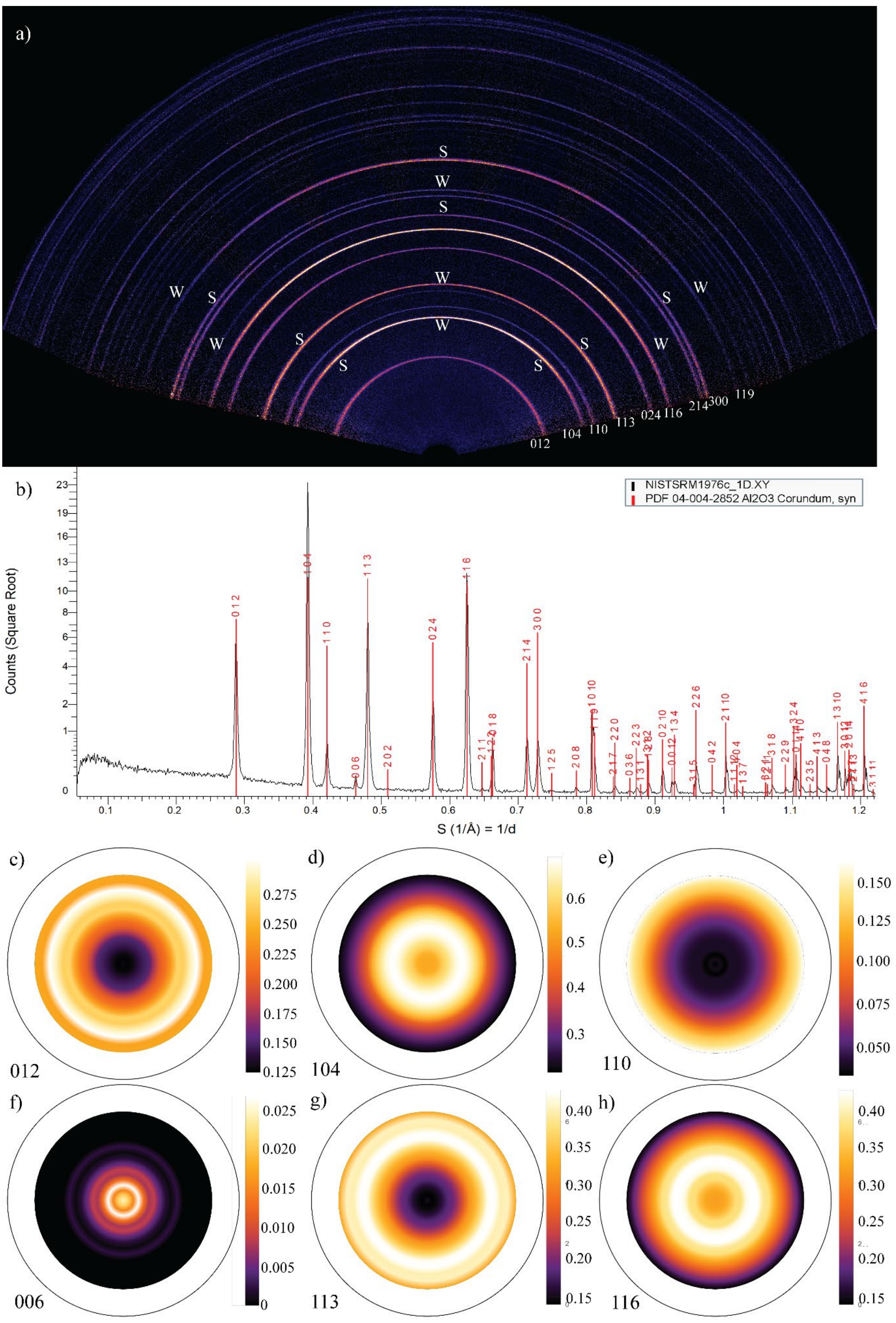

**Figure 9** **a)** WRRSM of NIST SRM 1976c corundum ceramic disk with main Debye arcs indexed. “S” denotes strong intensity parts while “W” denotes weak intensity parts of the same Debye arcs; **b)** Integrated 1D diffraction pattern shown in square root Y scale aligns with the ICDD PDF# 04-004-2852 of corundum phase; and the Pole Figures of **c)** 012, **d)** 104, **e)** 110, **f)** 006, **g)** 113, and **h)** 116 reflections expected from the corresponding integrated Debye arc profiles.

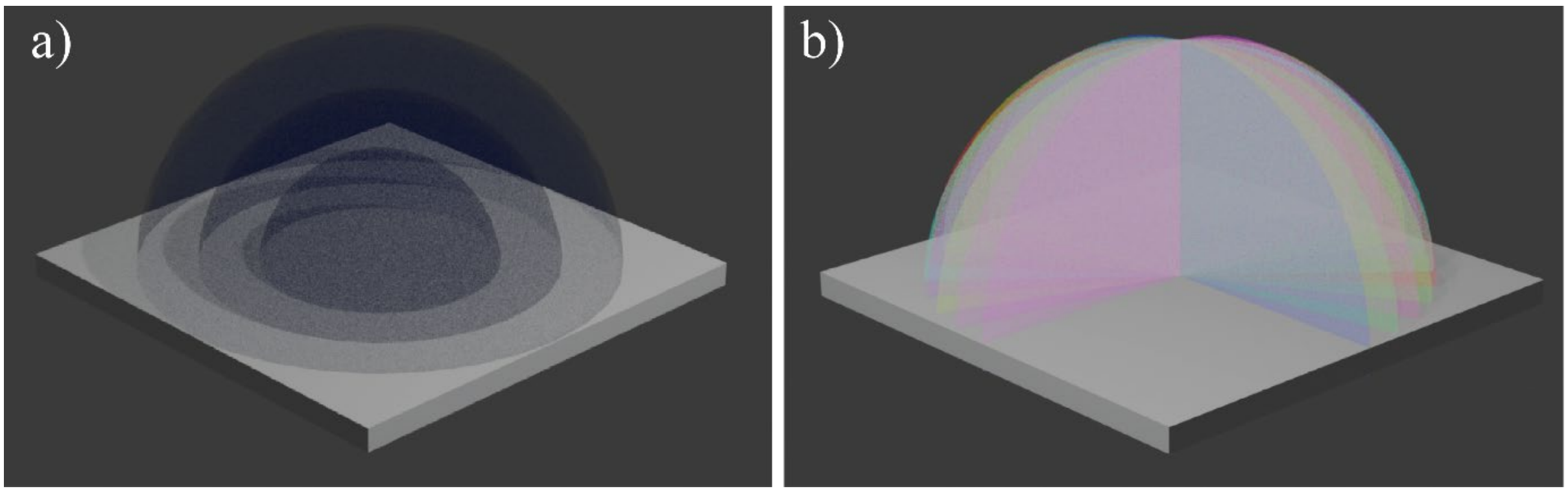

**Figure 10** **a)** The intensities of reciprocal space shells (pole figures) were scanned in conventional texture measurements; **b)** Same reciprocal space can be measured using multiple WRRSMs.

#### 6.4.2. Zenithal angle specific 1D diffraction pattern and residual stress estimation

Using the same integration function of the 1D diffraction pattern (Figure 9b) but for multiple sectors in different zenithal angle $\psi$ ranges, the lattice parameters of bulk samples in these zenithal angle directions can be refined from their corresponding 1D diffraction patterns. Figure 11 demonstrates simultaneous Pawley refinements of 1D diffraction patterns integrated from 7 different zenithal angle ranges evenly partitioned the full $-90° < \psi < 90°$ upper hemispheric directions on a WRRSM. The sample displacement parameters for corresponding symmetric $\psi$ ranges in both positive and negative directions were refined to a same value.

The values of lattice strain ε calculated from the variation of lattice parameters are plotted against $\mathrm{Sin}^2\psi$ values in Figure 12. The Young’s moduli and Poisson ratios for the two unit cell axes of corundum crystals are calculated from elastic constants of synthetic single crystal of corundum (Wachtman *et al.*, 1960):

$$\begin{aligned} E_a &= \frac{1}{S_{11}} = 425\text{GPa};\ \nu_a = -\frac{S_{12}}{S_{11}} = 0.304 \\ E_c &= \frac{1}{S_{33}} = 462\text{GPa};\ \nu_c = -\frac{S_{13}}{S_{33}} = 0.168 \end{aligned} \tag{19}$$

Compressive stresses derived from the “a” axis and “c” axis are 266MPa and 214MPa, respectively, based on the slopes of the ε – Sin$^2$ψ plots in Figure 12 (American Society for Testing and Materials, 2012):

$$\sigma_a = \frac{\text{slope}_a\ E_a}{(1+\nu_a)} = 266\ \text{MPa};\ \sigma_c = \frac{\text{slope}_c\ E_c}{(1+\nu_c)} = 214\ \text{MPa}; \tag{20}$$

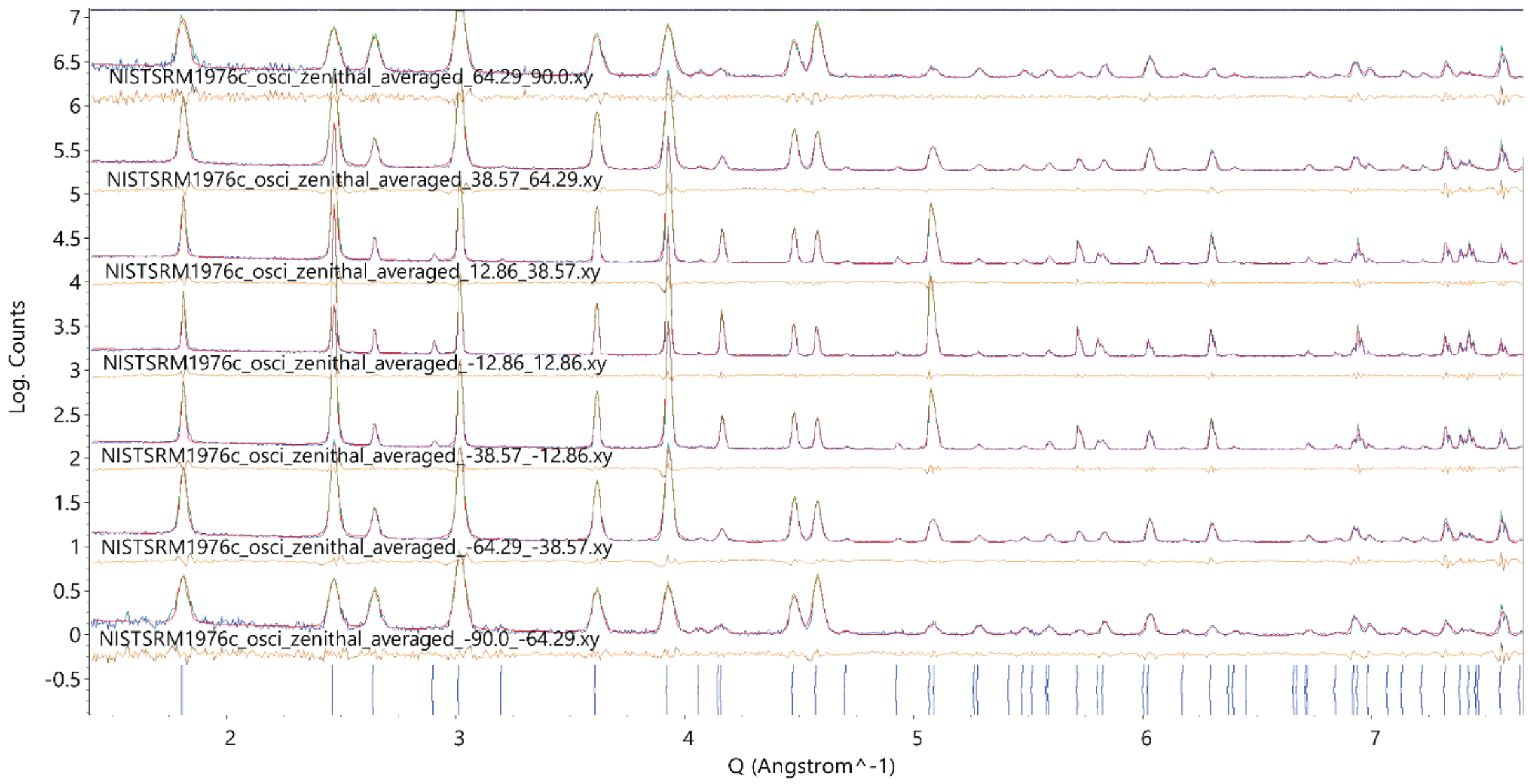

**Figure 11** Simultaneous Pawley fittings of 1D diffraction patterns from 7 different zenithal angle $\psi$ ranges: (-90.0° to -64.29°), (-64.29° to -38.57°), (-38.57° to -12.86°), (-12.86° to 12.86°), (12.86° to 38.57°), (38.57° to 64.29°), (64.29° to 90.0°) in DIFFRAC.TOPAS v7

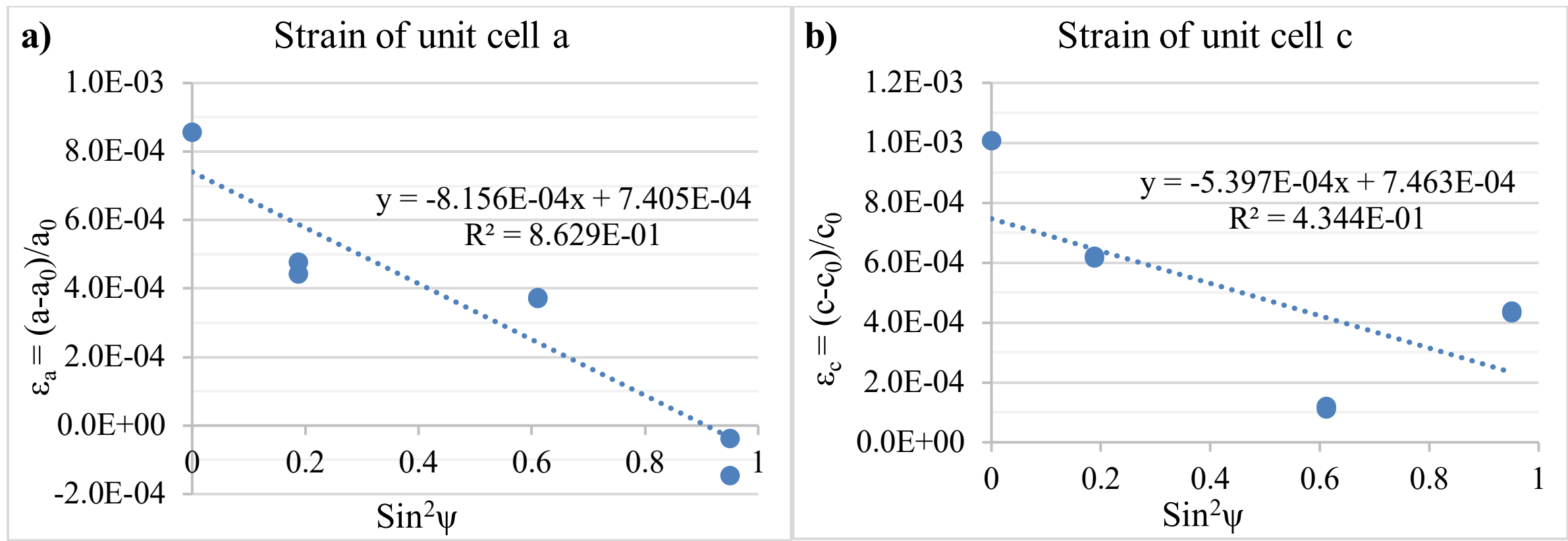

**Figure 12** Plots of strain ε values versus Sin$^2$ψ calculated from **a)** the refined lattice parameter “a”; **b)** the refined lattice parameter “c” from the Pawley fittings in Figure 11.

Comparing to the conventional residual stress measurement, which is normally derived from the peak position shifts of only a single *hkl* at high 2θ angle, the current proposed method from WRRSM is based on accurate lattice parameters refined from the whole 1D diffraction pattern integrated for the corresponding zenithal angle $\psi$ directions. Using this approach, it is possible to study anisotropic elasticity modulus and residual stresses in $\mathrm{Sin}^2\psi$ method for bulk materials.

### 7. Conclusion

The python program *RSstitcher* developed in this report merges 2D frames collected at different cradle tilt angles on laboratory X-ray diffractometers into Wide Range Reciprocal Space Maps (WRRSMs). The program demonstrated conversions of 2D frames in both symmetric scans for bulk samples and, for the first time, in ω-φ compensated SI-GID scans for thin film samples. WRRSM in the latter geometry provides similar information equivalent to GI-WAXS measurements on dedicated SAXS/WAXS instruments or on synchrotron beamlines. 2D frames collected on opposite azimuthal directions are merged into a full WRRSMs. *RSstitcher* outputs WRRSMs in quantitative tiff format which allows great flexibility in brightness and contrast control, annotation and labelling, as well as resolution customization using other common software, *e.g.* ImageJ. The customisable blurring option largely reduced the artifacts along the stitching edges of adjacent frames. *RSstitcher* is the first WRRSM software applied diffraction intensity correction due to flat sample self-absorption effect, hence provides intrinsic diffraction intensity links to the nature of the sample. This enables two classes of intensity integration functions in *RSstitcher*: 1) Exporting 1D diffraction patterns of equally divided zenithal wedges, the number of which are customisable by users; and 2) the intensity profiles along Debye arcs specifiable by users. The former function allows derivation of residual stress for bulk samples, and the latter function allows construction of pole figures for bulk samples. The webtool of *RSstitcher* running entirely on users' local browser makes this program installation-free and easy to use, without transferring any raw frame or result through internet.

**Acknowledgements** We acknowledge continued support from the Queensland University of Technology (QUT) through its centralisation and provision of subsidised access to specialised research infrastructure. We wish to acknowledge the support of the eResearch team at QUT for the provision of software engineering support, expertise, and research infrastructure in enablement of this project. The experimental data presented in this study were collected with the support from the Central Analytical Research Facility (CARF) at Queensland University of Technology (QUT). Dr. Peter Hines is thanked for the thickness measurement of PEDOT layer using an ellipsometer. A/Prof.

Christoph Schrank is thanked to provide the satin spar sample for data collection. Prof. Prashant Sonar is thanked to provide the PEDOT on PET sample for data collection.

**Conflicts of interest** The authors declare there are no conflicts of interest.

**Data availability** All data and source code are available on GitHub page: https://github.com/eresearchqut/RSstitcher. The webtool of *RSstiticher*: https://eresearchqut.github.io/RSstitcher/

## Appendix A. Analytical expression of ∠EOC in Figure 4a)

In Figure 4a), the length of |OE|, |OC|, and |EC| can be expressed in terms of |OB|:

$$|\mathrm{OE}| = \frac{|\mathrm{OB}|}{\cos(-\chi)} \tag{A1}$$

$$|\mathrm{OC}| = \frac{|\mathrm{OB}|}{\cos\theta'\cos\chi'} \tag{A2}$$

$$|\mathrm{EC}| = \sqrt{(|\mathrm{C'C}| + |\mathrm{EB}|)^2 + |\mathrm{C'B}|^2} = \sqrt{\left(\frac{|\mathrm{OB}|}{\cos\theta'\tan\chi'} + |\mathrm{OB}|\tan(-\chi)\right)^2 + (|\mathrm{OB}|\tan\theta')^2} \tag{A3}$$

The zenithal angle ∠EOC can then be derived from the law of cosines:

$$\angle\mathrm{EOC} = \cos^{-1}\left(\frac{|\mathrm{OE}|^2 + |\mathrm{OC}|^2 - |\mathrm{EC}|^2}{2|\mathrm{OE}|\cdot|\mathrm{OC}|}\right) = \cos^{-1}(\cos\chi\cos\theta'\cos\chi' + \sin\chi\sin\chi') \tag{A4}$$

# *RSstitcher*: Merging 2D diffraction frames for Wide Range Reciprocal Space Maps with absorption correction and integration functions

Authors

**Xiaodong Wang[a]*, Michael W. M. Jones[abcd] and Adam Smith[e]**

[a]Central Analytical Research Facility, Queensland University of Technology (QUT), Level 6, P Block, Gardens Point Campus, Brisbane, Queensland, 4001, Australia

[b]Centre for Materials Science, Queensland University of Technology, Brisbane, Queensland, 4001, Australia

[c]Planetary Surface Exploration Group, Queensland University of Technology, Brisbane, Queensland, 4001, Australia

[d]School of Chemistry & Physics, Queensland University of Technology, Brisbane, Queensland, 4001, Australia

[e]eResearch, Queensland University of Technology, Brisbane, Queensland, 4001, Australia

Correspondence email: tony.wang@qut.edu.au

# Supporting information

### S1. Problem of commercial WRRSM generation software

The WRRSM generated by commercial software “2D data Processor” (https://rigaku.com/service-and-support/software-download/2dp) and “3D Data Explorer” (https://rigaku.com/service-and-support/software-download/3d-explore) are compared with that from the current reported software *RSstitcher* in Figure S1. It is clear the high angle reciprocal spots highlighted in red circles were not merged correctly by this commercial software, while they were perfectly aligned by *RSstitcher* reported in this paper.

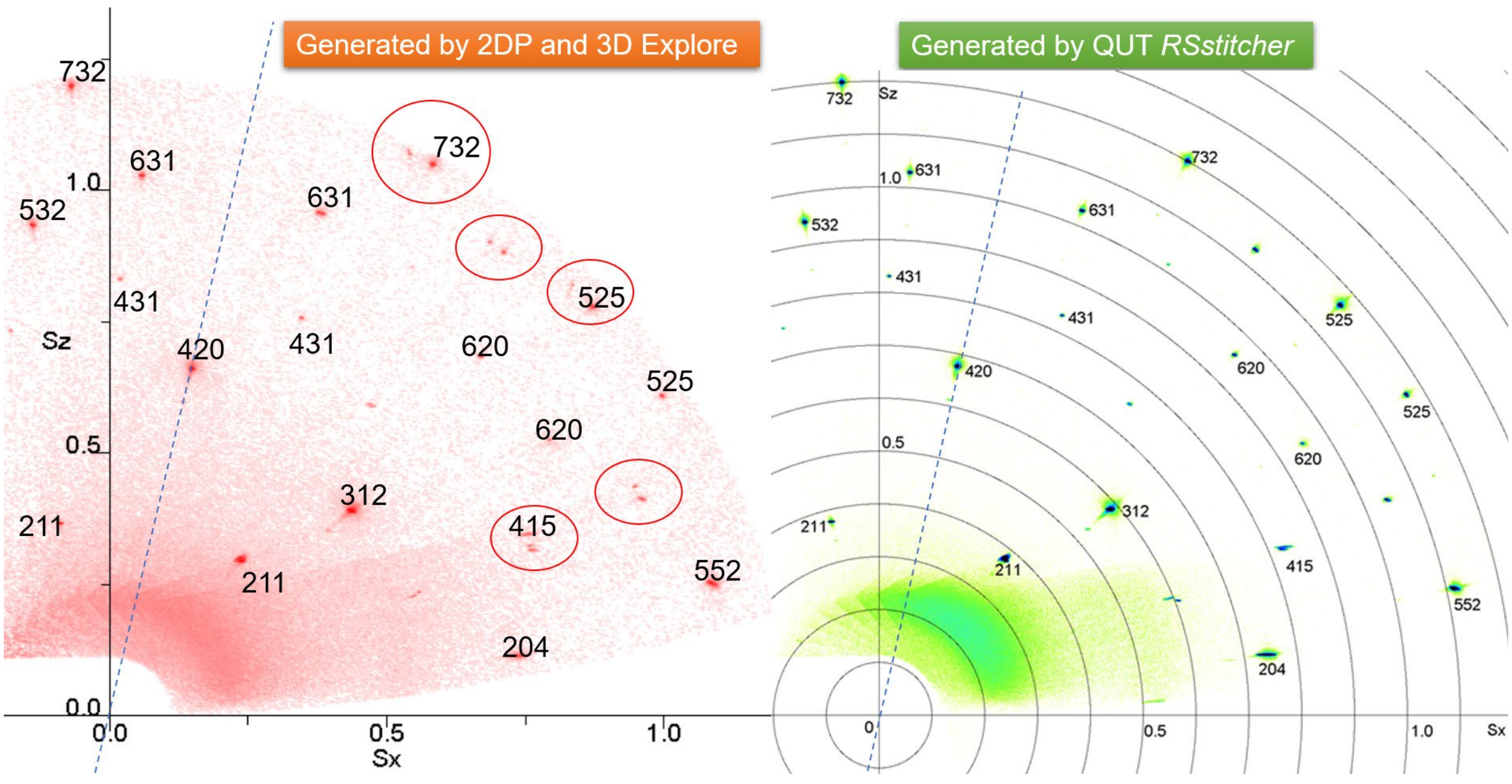

**Figure S1** Comparison of commercial software (left) and *RSstitcher* (right) on the same set of 2D frames collected in symmetric mode. The sample is a twined zircon crystal.

### S2. The Graphic User Interface (GUI) of *RSstitcher*

The screenshot of the full GUI of *RSstitcher* is shown in Figure S1. The top left section allows user to load their own data frames or choose an example data. The top right section allows user to adjust processing parameters, though usually not required. Typically, it takes the program less than 10 seconds to combine the loaded 2D frames into WRRSM, after user pressing the “Process” button. Once available, the calculated WRRSM will be shown immediately below the input section. The visualisation section of WRRSM allows overlaying a reciprocal space grid of $\Delta s = 0.1$ Å$^{-1}$.

An Experiment Parameters section summarising all the key experiment parameters extracted from the file head of the loaded 2D frames and output parameters. This information section also assists for debugging in case the output is not as user expected.

The “1D diffraction patterns” section plots integrated 1D powder patterns, the number of which matches with the number of zenithal bins user typed in. In below example, 3 zenithal bins were requested, therefore 3 1D patterns were generated from equally partitioned $\chi$ angle wedges (from -90 to -30 deg, from -30 to 30 deg, and from 30 to 90 deg).

An “Debye ring profiles” section forms the last output section, if user has crated Radial bins and assigned the $S_{min}$ to $S_{max}$ range for any particular Debye ring. In below example, 2 Debye rings were assigned, therefore their Debye ring profile is generated.

Each result section can be saved individual, or user may choose to save all results as a ZIP file by clicking “Download Results as ZIP” button next to the main “Process” button.

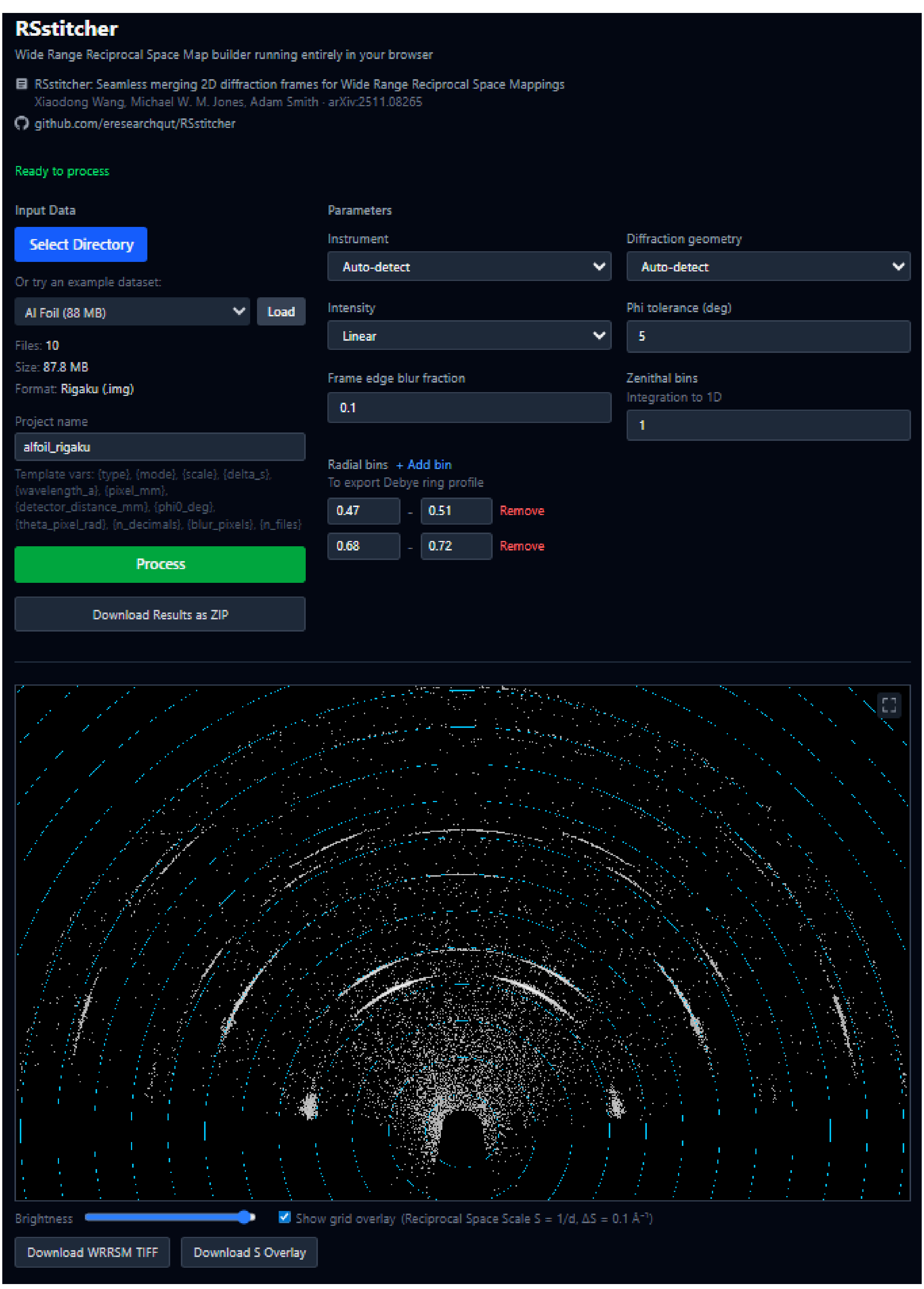

(a)

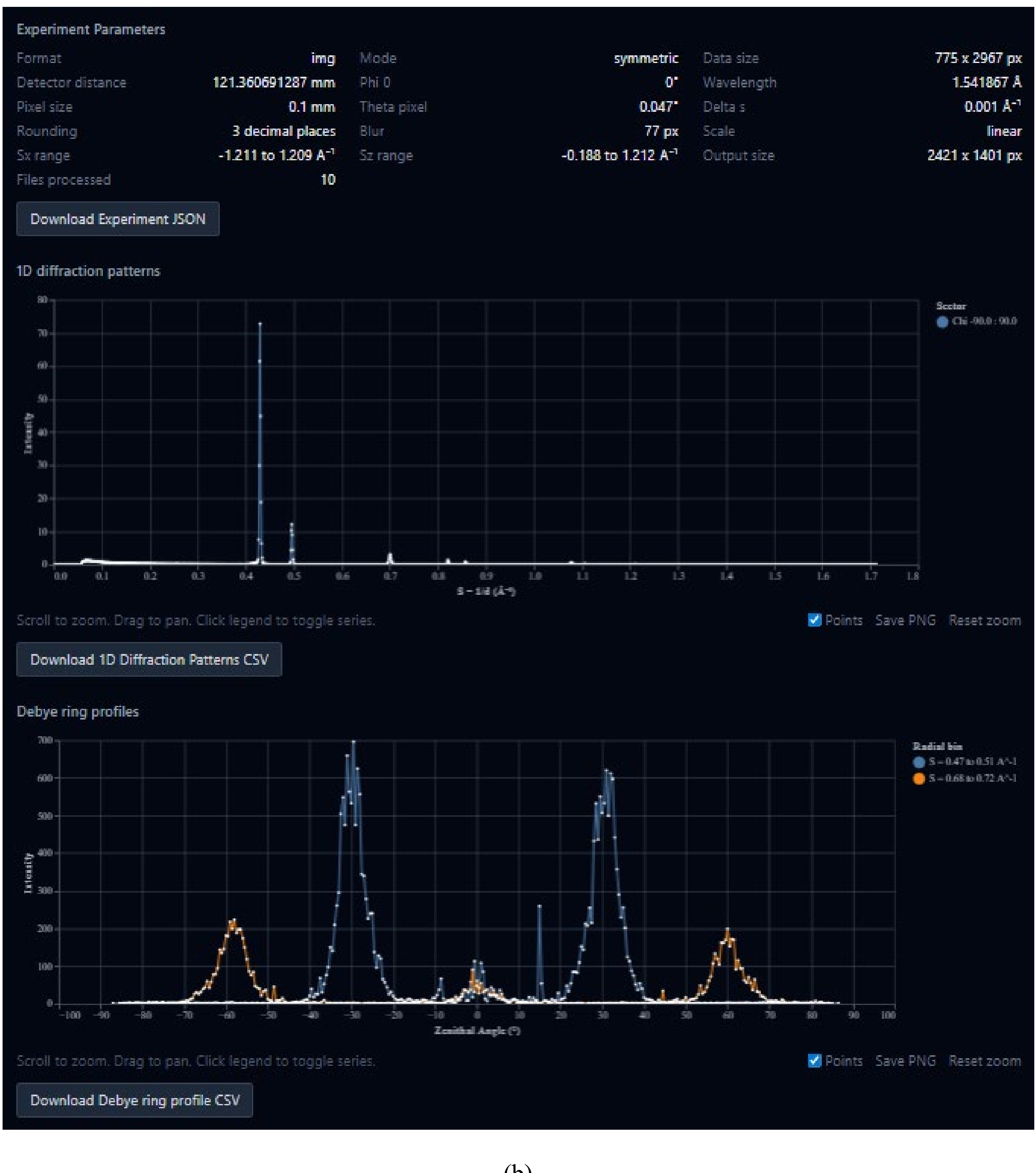

(b)

**Figure S2** The Graphic User Interface design of *RSstitcher* for ease of use. (a) *RSstitcher* allows mapping of reciprocal space up to $S_{max}$ ($1/d_{min}$) = 1.2 Å$^{-1}$, parameter settings, and integrated on-screen mouse measure of reciprocal space coordinates and intensities. (b) Quantitative analysis functions in *RSstitcher,* including integration into 1D powder diffraction data and intensity profile integration along Debye arcs, which allows subsequent pole figure generation and residual stress estimations.

### S3. WRRSM of NIST SRM 676a

As described in Section 6.1, the intensity correction devised in Eq.(13) recovers the intrinsic diffraction intensity of the nature of the sample. The WRRSM measured for a NIST SRM 676a corundum ceramic disk is shown in Figure S3. The reciprocal space circular ticks are omitted for clarity.

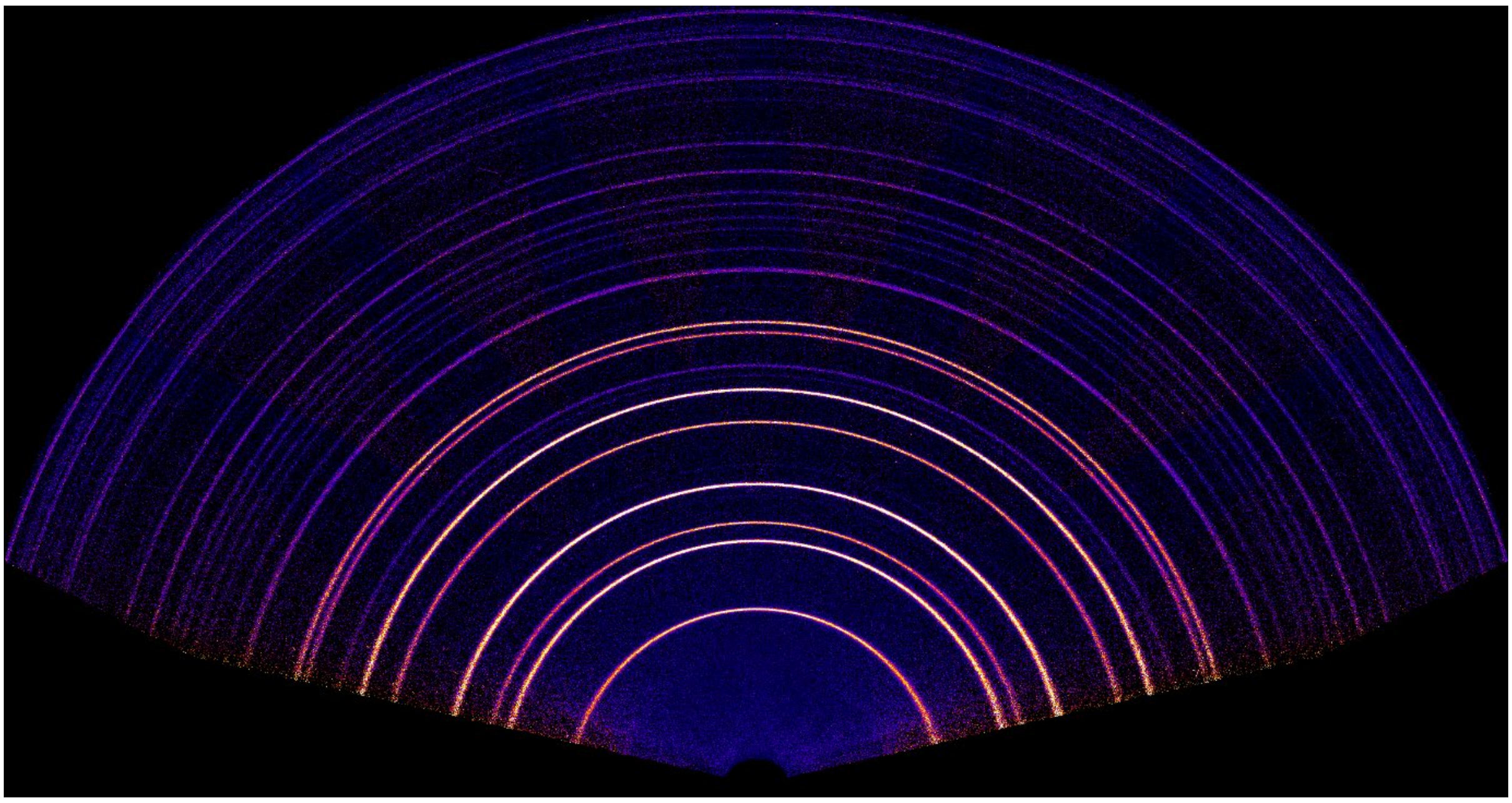

**Figure S3** *RSstitcher* generated WRRSM from 2D frames collected from a flattened surface of NIST SRM 676a corundum powder. Even intensity distributions along Debye arcs are achieved. The Debye arcs become broader at high $\psi$ angles due to longer X-ray footprint of the incident beam at high cradle tilt (defocusing effect).

### S4. The integrated intensity profiles of main Debye arcs of NIST SRM 1976c

The intensity profiles along Debye arcs exported from *RSstitcher* integrates an intensity of every $\Delta\chi_S = 1°$. Users can further re-bin the intensities according to the actual texture variations of the samples. Below Figure S4 were re-binned to every $\Delta\chi_S = 5°$ since the texture level of NIST SRM 1976c polycrystalline corundum ceramic disk is relatively weak.

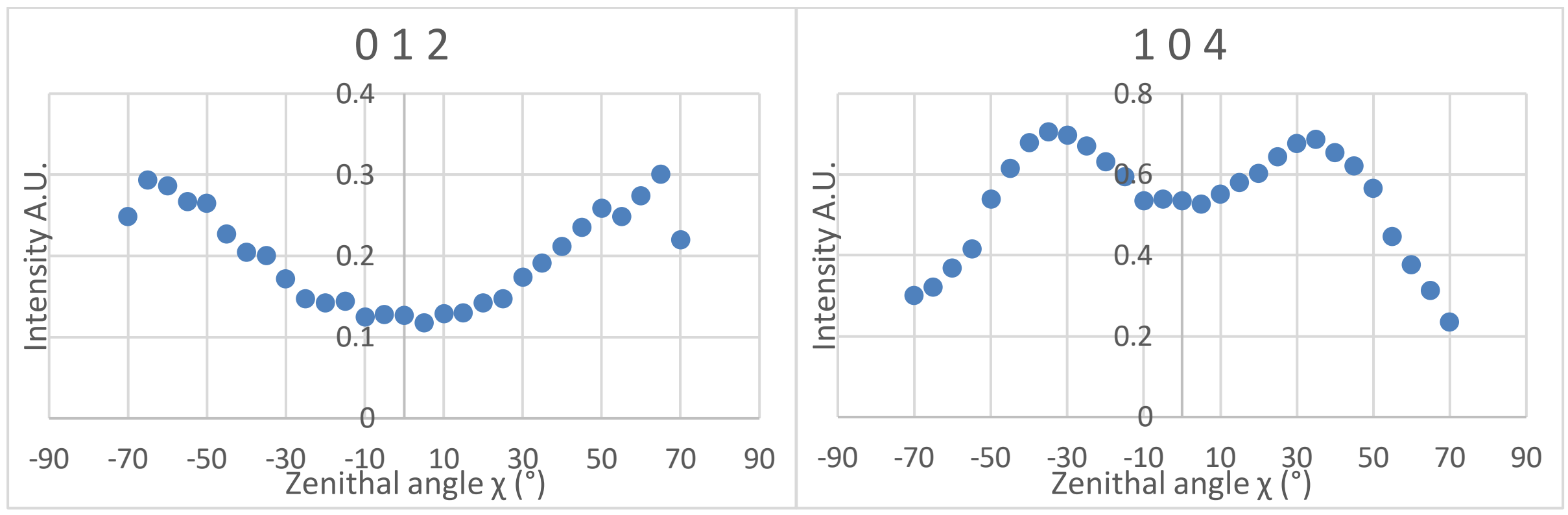

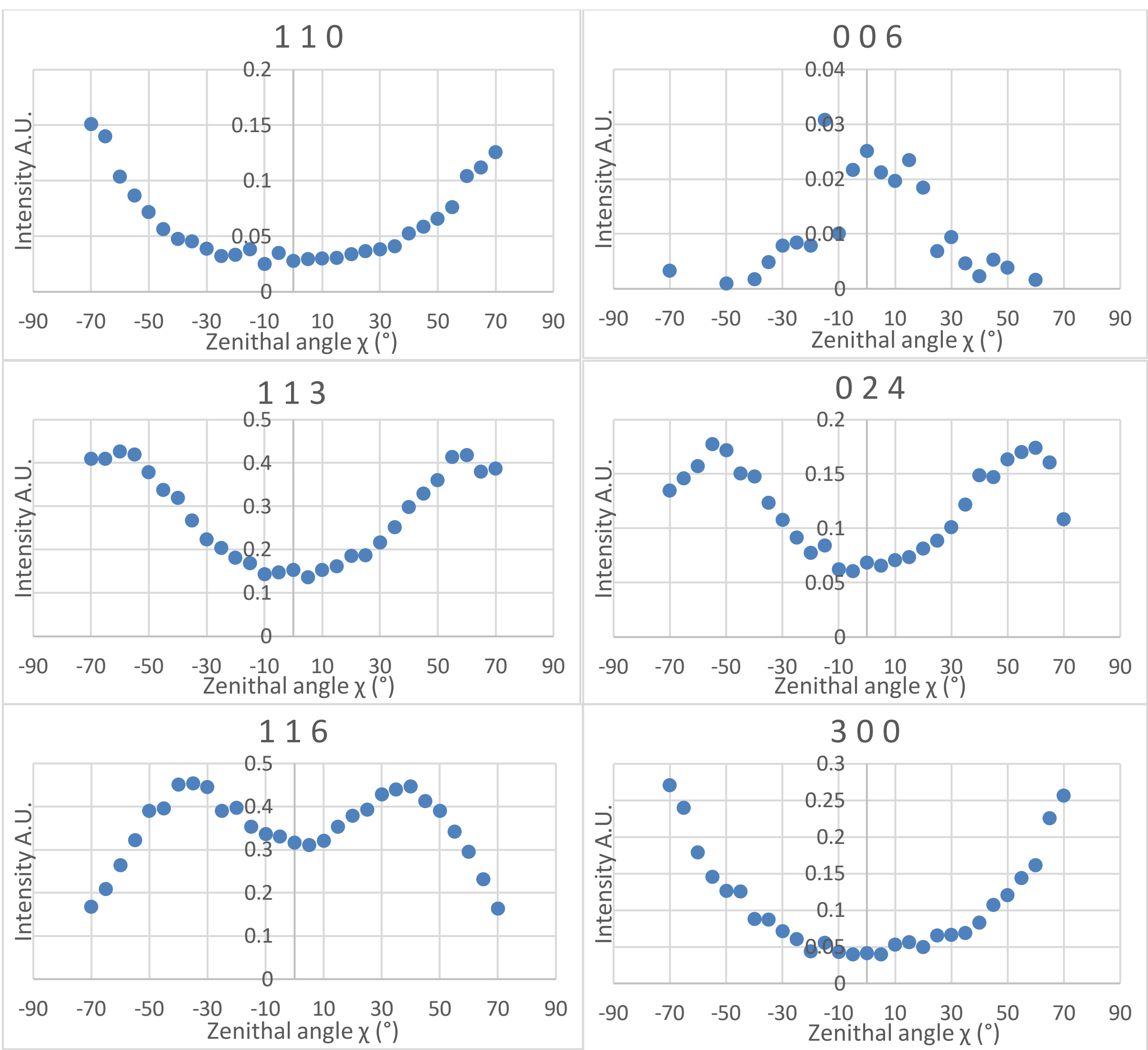

**Figure S4** *RSstitcher* exported intensity profiles on the corresponding *hkl* Debye arcs re-binned to every $\Delta\chi_S = 5°$ for the NIST SRM 1976c corundum ceramic disk.